\documentclass[useAMS,usenatbib]{mn2e}
\usepackage{lscape,graphicx,amssymb,aas_macros,bm}
\usepackage{amsmath} 
\usepackage{indentfirst} 
\usepackage{pdflscape} 
\usepackage{placeins} 
\usepackage{nicefrac} 
\usepackage{afterpage} 
\usepackage{booktabs} 
\usepackage{color} 
\usepackage{float}
\usepackage{subfloat}
\usepackage{adjustbox}

\newcommand{\lnhone}{\mbox{$\log\,N$(H\,{\scriptsize I})\ }}
\newcommand{\ewmgii}{\mbox{$W_r(2796)$\ }}
\newcommand{\ewciii}{\mbox{$W_r(977)$\ }}
\newcommand{\ewsiiii}{\mbox{$W_r(1206)$\ }}
\newcommand{\ewsiii}{\mbox{$W_r(1260)$\ }}
\newcommand{\ewovi}{\mbox{$W_r(1031)$\ }}
\newcommand{\ewlya}{\mbox{$W_r(1215)$\ }}
\newcommand{\apg}{\:^{>}_{\sim}\:}
\newcommand{\apl}{\:^{<}_{\sim}\:}
\newcommand{\cmjj}{\mbox{${\rm cm^{-2}}$}}
\newcommand{\etal}{et al.}
\newcommand{\HI}{{\mbox{H\,{\scriptsize I}}}}

\newcommand{\kms}{\mbox{km\ s${^{-1}}$}}
\newcommand{\lya}{\mbox{${\rm Ly}\alpha$}}
\newcommand{\lyb}{\mbox{${\rm Ly}\beta$}}

\newcommand{\AlII}{{\mbox{Al\,{\scriptsize II}}}}

\newcommand{\FeII}{{\mbox{Fe\,{\scriptsize II}}}}
\newcommand{\MgII}{{\mbox{Mg\,{\scriptsize II}}}}
\newcommand{\MgI}{{\mbox{Mg\,{\scriptsize I}}}}

\newcommand{\SiII}{{\mbox{Si\,{\scriptsize II}}}}
\newcommand{\CIII}{{\mbox{C\,{\scriptsize III}}}}
\newcommand{\CII}{{\mbox{C\,{\scriptsize II}}}}
\newcommand{\SiIII}{{\mbox{Si\,{\scriptsize III}}}}
\newcommand{\OVI}{{\mbox{O\,{\scriptsize VI}}}}

\newcommand{\mstar}{{\mbox{$M_{\rm star}$}}}
\newcommand{\msun}{{\mbox{M$_{\odot}$}}}

\title[CGM in Massive Quiescent Halos]{Characterizing Circumgalactic Gas around Massive Ellipticals at $\bm{z}\sim 0.4$ I. Initial Results\thanks{Based on data gathered with the 6.5m Magellan Telescopes located at Las Campanas Observatory, the W.~M.~Keck Observatory, and the NASA/ESA Hubble Space Telescope operated by the Space Telescope Science Institute and the Association of Universities for Research in Astronomy, Inc., under NASA contract NAS 5-26555.}}

\author[Chen et al.]{Hsiao-Wen Chen$^{1}$\thanks{E-mail: hchen@oddjob.uchicago.edu}, 
Fakhri S.\ Zahedy$^{1}$, Sean D.\ Johnson$^{2,3}$\thanks{Hubble \& Carnegie-Princeton Fellow}, Rebecca M.\ Pierce$^{1,4}$, \newauthor Yun-Hsin Huang$^{5}$, Benjamin J.\ Weiner$^{5}$, and Jean-Ren\'{e} Gauthier$^{6}$ \\ \\
$^{1}$Department of Astronomy \& Astrophysics, The University of Chicago, Chicago, IL 60637, USA \\
$^{2}$Department of Astrophysics, Princeton University, Princeton, NJ 08544, USA \\
$^{3}$The Observatories of the Carnegie Institution for Science, 813 Santa Barbara Street, Pasadena, CA 91101, USA \\
$^{4}$Department of Aerospace Engineering, University of Maryland, College Park, MD 20742, USA \\
$^{5}$Steward Observatory, University of Arizona, Tucson, AZ 85721, USA \\
$^{6}$DataScience.com, Culver City, CA 90230, USA \\
}

\begin{document}

\pagerange{\pageref{firstpage}--\pageref{lastpage}} \pubyear{2017}

\maketitle

\label{firstpage}

\begin{abstract}

We present a new {\it Hubble Space Telescope} ({\it HST}) Cosmic
Origins Spectrograph (COS) absorption-line survey to study 
halo gas around 16 luminous red galaxies (LRGs) at
$z=0.21-0.55$.  The LRGs are selected uniformly with stellar mass
$\mstar>10^{11}\,\msun$ and no prior knowledge of the presence/absence
of any absorption features.
%  The primary goals are (1) to constrain
%neutral hydrogen column density $N(\HI)$ based on observations of the
%full Lyman series and (2) to constrain the ionization state and
%chemical enrichment in massive quiescent halos based on observations
%of a suite of ionic transitions. 
Based on observations of the full Lyman series, we obtain
accurate measurements of neutral hydrogen column density $N(\HI)$ and
find that high-$N(\HI)$ gas is common in these massive quiescent halos
with a median of $\langle\,\log\,N(\HI)\rangle = 16.6$ at projected
distances $d\apl 160$ kpc.  We measure a mean covering fraction of
optically-thick gas with $\log\,N(\HI)\apg 17.2$ of
$\langle\kappa\rangle_{\rm LLS}=0.44^{+0.12}_{-0.11}$ at $d\apl 160$
kpc and $\langle\kappa\rangle_{\rm LLS}=0.71^{+0.11}_{-0.20}$ at
$d\apl 100$ kpc.  The line-of-sight velocity separations between the
\HI\ absorbing gas and LRGs are characterized by a mean and dispersion
of $\langle\,v_{{\rm gas}-{\rm gal}}\rangle=29$ \kms\ and
$\sigma_{\langle\,v_{{\rm gas}-{\rm gal}}\rangle}=171$ \kms.
Combining COS far-ultraviolet and ground-based echelle spectra provides an
expanded spectral coverage for multiple ionic transitions,
% including \SiII\,$\lambda\,1260$,
%\MgII\,$\lambda\lambda\,2796, 2803$, \CIII\,$\lambda\,977$,
%\SiIII\,$\lambda\,1206$, and \OVI\,$\lambda\lambda\,1031, 1037$.
from low-ionization \MgII\ and \SiII, to intermediate ionization
\SiIII\ and \CIII, and to high-ionization \OVI\ absorption lines.  We
find that intermediate ions probed by \CIII\ and \SiIII\ are the most
prominent UV metal lines in LRG halos with a mean covering fraction of
$\langle\kappa(\CIII)\rangle_{0.1}=0.75^{+0.08}_{-0.13}$ for
$\ewciii\ge 0.1$ \AA\ at $d<160$ kpc, comparable to what is seen for
\CIII\ in $L_*$ and sub-$L_*$ star-forming and red galaxies but
exceeding \MgII\ or \OVI\ in quiescent halos.  The COS-LRG survey
shows that massive quiescent halos contain widespread
chemically-enriched cool gas and that little distinction between LRG
and star-forming halos is found in their \HI\ and \CIII\ content.

\end{abstract}


\section{Introduction}
\label{section:introduction}

The circumgalactic medium (CGM), located in the space between galaxies
and the intergalactic medium (IGM), is regulated by the complex
interaction between IGM accretion and stellar feedback.  An accurate
characterization of the CGM is therefore critical for understanding
how galaxies grow and evolve.  Previous analytical calculations and
hydrodynamical simulations have suggested that chemically-pristine gas
accreted from the IGM by low-mass galaxies in less massive dark matter
halos ($M_{\rm halo}\apl 10^{12}\,M_\odot$) is cool and never
shock-heated to the virial temperature of the dark matter halo,
whereas more massive halos acquire their gas through a hot channel in
which the gas is shock-heated to temperatures $T\apg 10^6$ K (see
Faucher-Gigu\`ere 2017 for a recent review and a list of references).
At the same time, both cosmological simulations and observations of
the high-redshift Universe have also highlighted the importance of
supergalactic winds in enriching the CGM of star-forming galaxies and
the underdense IGM with heavy elements (see van de Voort 2017 for a
recent review and related references).  The presence of heavy elements
in turn alters the thermal state of the gas.

\begin{table*}
\scriptsize
\centering
\caption{Summary of the QSO--LRG Pair Sample}
\label{table:sample}
\centering {
\begin{tabular}{lrrclrrrrrr}
\hline \hline
\multicolumn{3}{c}{QSO} & & \multicolumn{7}{c}{LRG} \\
\cline{1-3} 
\cline{5-11}
 &   & \multicolumn{1}{c}{FUV} & & &  & \multicolumn{1}{c}{$\theta$} & \multicolumn{1}{c}{$d$} &  &  &  \\
\multicolumn{1}{c}{ID} &  \multicolumn{1}{c}{$z_{\rm QSO}$} & \multicolumn{1}{c}{(mag)} & &\multicolumn{1}{c}{ID} & \multicolumn{1}{c}{$z_{\rm LRG}$} & \multicolumn{1}{c}{(arcsec)} & \multicolumn{1}{c}{(kpc)} & \multicolumn{1}{c}{$(u-g)_{\rm rest}$} & \multicolumn{1}{c}{$M_r$} & \multicolumn{1}{c}{$\log\,M_{\rm star}/M_\odot^a$} \\
\hline 
SDSSJ\,094631.69$+$512339.9 & 0.741 & 18.4 & & SDSSJ\,094632.40$+$512335.9 & 0.4076 & 7.7 & 41.7 & $1.71\pm 0.12$ & $-22.42\pm 0.06$ & 11.2 \\
SDSSJ\,140626.60$+$250921.0 & 0.867 & 18.3 & & SDSSJ\,140625.97$+$250923.2 & 0.4004 & 8.8 & 47.3 & $1.77\pm 0.11$ & $-22.36\pm 0.04$ & 11.1 \\
SDSSJ\,111132.18$+$554726.1 & 0.766 & 17.5 & & SDSSJ\,111132.33$+$554712.8 & 0.4629 & 13.2 & 77.1 & $1.60\pm 0.08$ & $-23.15\pm 0.07$ & 11.4 \\
SDSSJ\,080359.23$+$433258.3 & 0.449 & 18.4 & & SDSSJ\,080357.74$+$433309.9$^b$ & 0.2535 & 19.9 & 78.5 & $1.69\pm 0.04$ & $-22.36\pm 0.02$ & 11.1 \\
SDSSJ\,092554.70$+$400414.1 & 0.471 & 18.2 & & SDSSJ\,092554.18$+$400353.4$^b$ & 0.2475 & 21.6 & 83.7 & $1.82\pm 0.04$ & $-22.28\pm 0.02$ & 11.1 \\
SDSSJ\,095000.73$+$483129.2 & 0.590 & 17.9 & & SDSSJ\,095000.86$+$483102.2$^b$ & 0.2119 & 27.1 & 93.7 & $1.64\pm 0.03$ & $-22.27\pm 0.02$ & 11.0 \\
SDSSJ\,112756.76$+$115427.1 & 0.509 & 18.1 & & SDSSJ\,112755.83$+$115438.3 & 0.4237 & 17.7 & 98.7 & $1.93\pm 0.16$ & $-22.45\pm 0.06$ & 11.2 \\
SDSSJ\,124307.57$+$353907.1 & 0.547 & 18.4 & & SDSSJ\,124307.36$+$353926.3 & 0.3896 & 19.3 & 102.2 & $1.90\pm 0.11$ & $-22.81\pm 0.04$ & 11.3 \\
SDSSJ\,155048.29$+$400144.9 & 0.497 & 18.1 & & SDSSJ\,155047.70$+$400122.6$^b$ & 0.3125 & 23.3 & 106.7 & $1.54\pm 0.05$ & $-22.51\pm 0.02$ & 11.2 \\
SDSSJ\,024651.91$-$005930.9 & 0.468 & 18.0 & & SDSSJ\,024651.20$-$005914.1 & 0.4105 & 19.9 & 108.6 & $1.75\pm 0.07$ & $-23.02\pm 0.03$ & 11.4 \\
SDSSJ\,135726.27$+$043541.4 & 1.233 & 19.2 & & SDSSJ\,135727.27$+$043603.3 & 0.3296 & 26.5 & 125.9 & $1.77\pm 0.07$ & $-22.71\pm 0.03$ & 11.3 \\
SDSSJ\,091029.75$+$101413.5 & 0.463 & 18.5 & & SDSSJ\,091027.70$+$101357.2$^b$ & 0.2641 & 34.4 & 140.1 & $1.77\pm 0.04$ & $-22.69\pm 0.02$ & 11.2 \\
SDSSJ\,141309.14$+$092011.2 & 0.460 & 17.5 & & SDSSJ\,141307.39$+$091956.7 & 0.3584 & 29.8 & 149.2 & $1.70\pm 0.05$ & $-23.83\pm 0.02$ & 11.7 \\
SDSSJ\,155304.92$+$354828.6 & 0.721 & 17.7 & & SDSSJ\,155304.32$+$354853.9 & 0.4736 & 26.3 & 155.9 & $1.54\pm 0.11$ & $-22.15\pm 0.11$ & 11.0 \\
SDSSJ\,125901.67$+$413055.8 & 0.745 & 18.4 & & SDSSJ\,125859.98$+$413128.2 & 0.2790 & 37.6 & 159.1 & $1.71\pm 0.03$ & $-23.49\pm 0.02$ & 11.6 \\
SDSSJ\,124410.82$+$172104.6 & 1.282 & 18.4 & & SDSSJ\,124409.17$+$172111.9 & 0.5591 & 24.8 & 160.1 & $1.73\pm 0.06$ & $-23.35\pm 0.09$ & 11.5 \\
\hline
%		\hline \multicolumn{12}{l}{\bf Notes} \\
\multicolumn{11}{l}{$^\mathrm{a}$Uncertainties in $M_{\rm star}$ are known to be better than 0.2 dex (e.g., Conroy 2013).} \\
\multicolumn{11}{l}{$^\mathrm{b}$COS-Halos red galaxies that match our LRG selection criteria; projected distances updated based on our own calculations.} \\
\end{tabular}
}
\end{table*}

Over the last decade, substantial effort has been made to understand
the physical mechanisms driving the evolution of the CGM (see Chen
2017a and Tumlinson \etal\ 2017 for recent reviews), but the origin of
the cool, $T \sim 10^{4-5}$ K circumgalactic gas is largely uncertain.
In particular, the observed high incidence of chemically-enriched cool
gas in the vicinities of luminous red galaxies (LRGs; e.g., Gauthier
\etal\ 2009, 2010, Lundgren \etal\ 2009; Gauthier \& Chen 2011; Bowen
\& Chelouche 2011; Zhu \etal\ 2014; Huang \etal 2016) remains a
puzzle, because LRGs are massive elliptical galaxies at $z\sim 0.5$
and cool gas is not expected to survive in their host halos.

By design, the LRGs are selected in the Sloan Digital Sky Survey
(SDSS; York \etal\ 2000) to exhibit colors that resemble nearby
elliptical galaxies (Eisenstein et al.\ 2001).  They are characterized
by luminosities of $\apg 3\,L_*$ and stellar masses of $\mstar\,\apg
10^{11} M_\odot$ at $z\approx 0.5$ (e.g., Tojeiro et al.\ 2011), and
exhibit little on-going or recent star formation (e.g., Roseboom
\etal\ 2006; Gauthier \& Chen 2011; Huang \etal\ 2016).  The observed
strong clustering amplitude of these galaxies indicates that they
reside in halos of $M_{\rm halo} \apg 10^{13}\,\msun$ (e.g., Zheng et
al.\ 2007; Blake et al.\ 2008; Padmanabhan et al.\ 2007).  Independent
studies have also shown that more than 90\% of massive galaxies with
$\mstar \apg 10^{11}\,\msun$ in the local universe contain primarily
evolved stellar populations (e.g., Peng et al.\ 2010; Tinker et
al.\ 2013), making the LRGs an ideal laboratory for studying the cold
gas content in massive quiescent halos.

Gauthier et al.\ (2009) noted that the cross-correlation signal of
photometrically selected LRGs and \MgII\ absorbers ($z\sim 0.5$) is
comparable to the LRG auto-correlation on scales of $r_p\apl 400$ kpc.
The comparable clustering amplitudes on scales of the LRG halo size
suggests that a large fraction of the LRGs host a \MgII\ absorber,
which is understood to originate in photo-ionized gas of temperature
$T \sim 10^4$ K (Bergeron \& Stas\'inska 1986; Charlton et al.\ 2003)
and trace high-column density clouds of neutral hydrogen column
density $N(\HI) \approx 10^{16}-10^{22}$ \cmjj\ (Rao et al.\ 2006).
Subsequent spectroscopic studies have indeed uncovered strong Mg\,II
absorbers in a significant fraction of LRG halos with a mean covering
fraction of $\kappa({\MgII})\apg 15\,(5)$\% at projected distances $ d<
120\,(500)$ kpc (Gauthier \etal\ 2010; Huang \etal\ 2016).  In
addition, strong \lya\ absorption is found common around early-type
galaxies at $\langle\,z\,\rangle\approx 0.2$ (Thom \etal\ 2012;
Tumlinson \etal\ 2013).  Furthermore, utilizing multiply-lensed QSOs,
Zahedy \etal\ (2016, 2017a) examined the gas content of the lensing
galaxies, which are also massive ellipticals at $z=0.4-0.7$, and found
that high column density gas exists at projected distances as small as
$\approx 3-15$ kpc from these ellipticals.  While absorption-line
studies continue to show the presence of chemically-enriched cool gas
around red galaxies at intermediate redshifts, local 21~cm and CO
surveys of elliptical galaxies have also uncovered cold atomic and/or
molecular gas in $>30$\% of nearby ellipticals (e.g., Oosterloo
\etal\ 2010; Serra \etal\ 2012; Young \etal\ 2014).  Together, these
independent studies show that cool gas outlasts star formation in
massive quiescent halos over an extended cosmic time period.

The presence of cool gas in massive halos challenges theoretical
expectations of gaseous halos from both simple analytic models and
numerical simulations.  The predominantly old stellar population and
quiescent state in these passive galaxies also make it difficult to
apply feedback due to starburst and active galactic nuclei (AGN) as a
general explanation for the presence of these absorbers (e.g., Rahmati
\etal\ 2015; Faucher-Gigu\`ere \etal\ 2016).
%The presence of cool gas in massive quiescent halos has important
%implications for the baryon content of individual galactic halos,
%including the total baryonic mass in the cool phase that has eluded
%detections in previous searches and the state of the confining hot
%halos.  However, it has been difficult to 
Applying the large number of SDSS \MgII\ absorption observations of
LRGs at intermediate redshifts to test theoretical predictions
requires knowledge of the ionization state and metallicity of the gas,
which are not available in most cases.  To facilitate direct and
quantitative comparisons between observations and theoretical
predictions, we are carrying out a systematic study of the CGM around
LRGs using the Cosmic Origins Spectrograph (COS) on board the {\it
  Hubble Space Telescope} ({\it HST}).  The spectral coverage of COS
enables observations of the full H\,I Lyman series transitions and
metal absorption features that probe halo gas under different
ionization conditions, including \CIII\,$\lambda\,977$,
\OVI\,$\lambda\lambda\,1031, 1037$, \SiIII\,$\lambda\,1206$, and
\SiII\,$\lambda\,1260$.  The LRGs in our sample are selected based
only on their proximity to the sightline of a UV bright background
QSO.  No prior knowledge of the halo gas properties was used in the
selection.  Here we present initial results from our program.

The paper is organized as follows.  In Section 2, we describe the
design of the COS-LRG program and related spectroscopic observations.
In Section 3, we present absorption-line measurements.  In Section 4,
we discuss the general properties found for LRG halos and their
implications.  We adopt a standard $\Lambda$ cosmology, $\Omega_M$ =
0.3 and $\Omega_\Lambda$=0.7 with a Hubble constant $H_{\rm 0} = 70\rm
\,km\,s^{-1}\,Mpc^{-1}$ throughout the paper.

\section{The COS-LRG Program}
\label{section:data}

To advance a deeper understanding of gas properties in massive
quiescent halos, we are carrying out a COS-LRG program for a
systematic study of the CGM around LRGs at intermediate redshifts.
The primary goals of the program are (1) to obtain accurate and
precise measurements of neutral hydrogen column density $N(\HI)$ based
on observations of the full hydrogen Lyman series and (2) to constrain
the ionization state and chemical enrichment in massive quiescent
halos based on observations of a suite of ionic transitions.  The LRGs
in our sample are selected without prior knowledge of the
presence/absence of any absorption features in the halos and therefore
enable an accurate assessment of the chemical enrichment level in
these massive quiescent halos.  Here we describe our program design
and associated spectroscopic data for the subsequent absorption-line
studies.

\subsection{Program Design}

To establish a uniform sample of LRGs for a comprehensive study of the
ionization state and chemical enrichment in massive quiescent halos,
we cross-correlated UV-bright quasars with ${\rm FUV}\apl 18.5$ mag
and all LRGs found at $z\apg 0.26$ in the literature.  We identified
spectroscopically confirmed LRGs in the SDSS archive that occur at
$d\apl 160$ kpc from the sightline of a UV-bright background quasar to
ensure that high-resolution ($R\equiv\lambda/\Delta\,\lambda\approx
18000$) FUV absorption spectra of the QSOs can be obtained using the
Cosmic Origins Spectrograph (COS) on board the {\it Hubble Space
  Telescope} ({\it HST}) for observing and resolving weak ionic
absorption features.  The maximum projected distance $d=160$ kpc
corresponds to roughly $1/3$ of the virial radius ($R_{\rm vir}$) of a
$10^{13}\,\msun$ dark matter halo.  It was motivated by the SDSS
\MgII\ survey of Huang \etal\ (2016), who found a mean covering
fraction of $>10$\% at $d\le 160$ kpc around LRGs for $\ewmgii > 0.3$
\AA\ absorbers and $>70$\% for weaker ones with $\ewmgii > 0.1$ \AA.
We limited our search to LRGs at $z\apg 0.26$ to ensure the ability to
constrain neutral hydrogen column density, $N({\HI})$, from
observations of higher-order Lyman absorption series.  In addition, we
restricted our search to LRGs that occur at a line-of-sight velocity
difference of $\Delta v < -10,000$ \kms\ from the background QSO, to
avoid confusion with outflowing gas from the quasar (e.g., Wild
\etal\ 2008).

This exercise yielded a unique sample of 16 LRGs with a UV-bright
background QSO found at $d\apl 160$ kpc.  Because both UV bright QSOs
and LRGs are rare, the number of projected LRG-QSO pairs with close
separations is small.  This modest sample size underscores the
challenge of probing the CGM in massive quiescent halos using
absorption-line techniques.  A summary of the LRG--QSO pair sample is
presented in Table 1, which lists for each pair the QSO ID, redshift
($z_{\rm QSO}$), and FUV magnitude observed by GALEX in the first
three columns, and the LRG ID, redshift ($z_{\rm LRG}$), angular
distance ($\theta$) and the corresponding physical projected distance
($d$) to the QSO sightline, the rest-frame $u-g$ color, rest-frame
$r$-band magnitude ($M_r$), and total stellar mass ($M_{\rm star}$) in
the following seven columns.  For each LRG, the rest-frame $u-g$ color
and $M_r$ were estimated by interpolating between available optical
broad-band photometry in the observed SDSS $u$, $g$, $r$, $i$, and $z$
bandpasses, and $M_{\rm star}$ was estimated using the $K$-correct
code (Blanton \& Roweis 2007) for a Chabrier Initial Mass Function
(Chabrier 2003).  Error bars in $u-g$ and $M_r$ reflect 1-$\sigma$
uncertainties in the photometric measurements.  Uncertainties in
$M_{\rm star}$ are understood to be dominated by systematic
differences in the model priors and are known to be less than 0.2 dex
(e.g., Moustakas \etal\ 2013; Conroy 2013).

\begin{figure}
\includegraphics[scale=0.43]{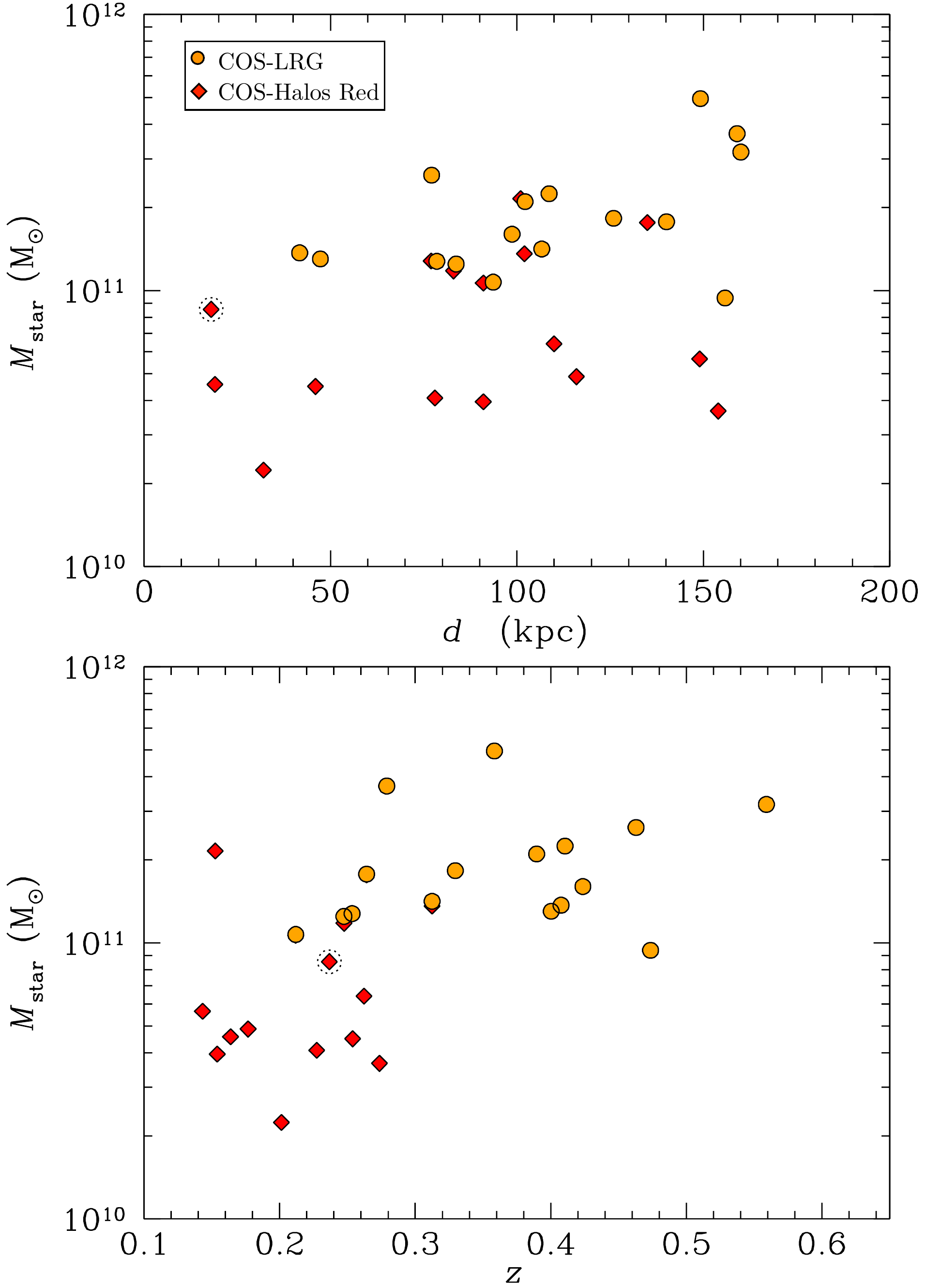}
\caption{Comparisons of the LRGs in our studies (orange filled
  circles) and the red galaxies in the COS-Halos sample (red diamonds)
  considered by Thom \etal\ (2012).  The LRGs form a unique,
  mass-limited sample of quiescent galaxies located at $d\apl 160$ kpc
  from a UV bright background QSO.  \mstar\ of the LRGs ranges from
  $\log\,\mstar=11$ to $\log\,\mstar\,/\,\msun\,=\,11.7$ with a median
  of $\langle\,\mstar\,/\,\msun\,\rangle_{\rm COS-LRG}=11.2$.  In
  contrast, the COS-Halos red galaxies span a wide range in
  \mstar\ from $\log\,\mstar\,/\,\msun\,=\,10.3$ to
  $\log\,\mstar\,/\,\msun\,=\,11.3$ with a median of
  $\langle\,\mstar\,/\,\msun\,\rangle_{\rm COS-Halos Red}=10.8$.  Note
  that one of the COS-Halos red galaxies, J1133$+$032\_110\_5 at
  $z=0.237$ and $d=18$ kpc (red diamond with a dotted circle), has a
  neighboring QSO at the same redshift and $d=124$ kpc.  Five of the
  remaining COS-Halos red galaxies satisfy the LRG selection criteria,
  and are therefore included in the LRG sample.  These are shown in
  both sets of symbols.  We have re-calculated both \mstar\ and $d$
  ourselves for consistency, which led to small offsets in $d$.}
\label{figure:comparison}
\end{figure}

Five of the LRGs in our sample overlap with the COS-Halos red galaxy
subsample (e.g., Tumlinson \etal\ 2011; Thom \etal\ 2012).
Comparisons of our COS-LRG sample with the red galaxies studied by the
COS-Halos team (e.g., Thom \etal\ 2012) are presented in Figure 1,
which displays the $M_{\rm star}$ versus $d$ distribution in the top
panel and the $M_{\rm star}$ versus $z$ distribution in the bottom
panel.  Note that we have re-calculated \mstar\ for all COS-Halos
galaxies based on available SDSS photometric measurements.  Our
estimates of \mstar\ using $K$-correct are found to be systematically
lower than those reported in Werk \etal\ (2012) by 0.2 dex.  The
discrepancy can be attributed to a different stellar initial mass
functions adopted by these authors.  For consistency, we adopt our
\mstar\ estimates for all COS-Halos galaxies in comparisons with the
COS-LRG sample.  In addition, one galaxy in the COS-Halos red galaxy
sample (J1617$+$0638\_253\_39) satisfies our mass selected criterion
but is not included in the COS-LRG sample.  This red galaxy occurs at
$d=101$ kpc from the background QSO sightline and has $\mstar\approx
2\times 10^{11}\,\msun$, but the redshift is too low that the Lyman
series transitions are not covered by available COS spectra.  While
\lya\ is not detected with a 2-$\sigma$ upper limit to the rest-frame
absorption equivalent width of $W_r(1215)=0.14$ \AA\ (Thom
\etal\ 2012), we exclude this from the COS-LRG in order not to bias
subsequent analysis on the incidence of cool gas in these massive
halos.  We also note that one of the COS-Halos red galaxies,
J1133$+$032\_110\_5 at $z=0.237$ and $d=18$ kpc (red diamond with a
dotted circle in Figure 1), has a neighboring QSO at the same redshift
and $d=124$ kpc.  Quasar host halos have been shown to exhibit CGM
properties that are different from both starburst and passive galaxies
(e.g., Prochaska \etal\ 2014; Johnson \etal\ 2015a; see also Chen
2017b for relevant discussions).

Figure 1 demonstrates that the LRGs form a uniform sample of high-mass
halos with \mstar\ ranging from $\log\,\mstar\,/\,\msun\,=\,11$ to
$\log\,\mstar\,/\,\msun\,=\,11.7$ and a median of
$\langle\,\log\,\mstar\,/\,\msun\,\rangle_{\rm COS-LRG}=11.2$ over the
redshift range from $z=0.21$ to $z=0.55$.  In contrast, the COS-Halos
red galaxy sample includes predominantly lower-mass halos with
\mstar\ ranging from $\log\,\mstar\,/\,\msun\,=\,10.3$ to
$\log\,\mstar\,/\,\msun\,=\,11.3$ and a median of
$\langle\,\log\,\mstar\,/\,\msun\,\rangle_{\rm COS-Halos Red}=10.8$
over the redshift range from $z=0.14$ to $z=0.27$.  The COS-LRG sample
therefore offers a unique opportunity both for studying the CGM in the
most massive individual galaxy halos and for comparison with less
massive but still quiescent systems.

\begin{table}
\scriptsize
\centering
\caption{Journal of {\it HST} COS Observations}
\label{table:cosobs}
\centering {
\begin{tabular}{lrrr}
\hline \hline
\multicolumn{1}{c}{QSO} & \multicolumn{1}{c}{Grating} & \multicolumn{1}{c}{$t_{\rm exp}$ (s)} & \multicolumn{1}{c}{PID} \\
\hline 
SDSSJ\,024651.91$-$005930.9 & G130M/G160M & 5121/11129 & 14145 \\
SDSSJ\,080359.23$+$433258.3 & G130M/G160M & 5207/6110 & 11598 \\
SDSSJ\,091029.75$+$101413.5 & G130M/G160M & 4913/8700 & 11598 \\
SDSSJ\,092554.70$+$400414.1 & G130M/G160M & 3765/4303 & 11598 \\
SDSSJ\,094631.69$+$512339.9 & G130M/G160M & 5655/15166 & 14145 \\
SDSSJ\,095000.73$+$483129.2 & G130M/G160M & 2445/2927 & 11598 \\
                            & G130M & 2488 & 13033 \\
SDSSJ\,111132.18$+$554726.1 & G130M/G160M & 8388/8850 & 12025 \\
SDSSJ\,112756.76$+$115427.1 & G130M/G160M & 5147/11173 & 14145 \\
SDSSJ\,124307.57$+$353907.1 & G130M/G160M & 10446/20286 & 14145 \\
SDSSJ\,124410.82$+$172104.6 & G160M & 7172 & 12466 \\
SDSSJ\,125901.67$+$413055.8 & G130M/G160M & 8230/14223 & 13833 \\
SDSSJ\,135726.27$+$043541.4 & G130M/G160M & 14148/28205 & 12264 \\
SDSSJ\,140626.60$+$250921.0 & G130M/G160M & 5195/14184 & 14145 \\
SDSSJ\,141309.14$+$092011.2 & G130M/G160M & 4959/8138 & 13833 \\
SDSSJ\,155048.29$+$400144.9 & G130M/G160M & 3804/4160 & 11598 \\
SDSSJ\,155304.92$+$354828.6 & G130M/G160M & 2114/2837 & 11598 \\
\hline
%		\hline \multicolumn{12}{l}{\bf Notes} \\
%\multicolumn{11}{l}{$^\mathrm{a}$COS-Halos red galaxies that match our LRG selection criteria.} \\
\end{tabular}
}
\end{table}

\subsection{COS UV Spectroscopy}

High-resolution FUV absorption spectra of the background QSOs are
available either from our own COS program (PID$\,=\,$14145) or in the
{\it HST} COS data archive.  COS with the G130M and G160M gratings and
different central wavelengths offers a contiguous spectral coverage
from $\lambda\approx 1140$ \AA\ to $\lambda\approx 1780$ \AA, with a
spectral resolution of Full-Width-at-Half-Maximum ${\rm FWHM}\approx
17$ \kms.  The spectral coverage of COS enables observations of the
full H\,I Lyman series transitions and metal absorption features that
probe halo gas under different ionization conditions.  Common metal
absorption lines observable by COS include \CIII\,$\lambda\,977$,
\OVI\,$\lambda\lambda\,1031, 1037$, \SiIII\,$\lambda\,1206$, and
\SiII\,$\lambda\,1260$.  A summary of the COS observations is
presented in Table 2, which lists for each QSO the grating used for
the spectra, the total exposure time per grating in seconds, and the
program ID under which the data were obtained.  Table 2 shows that all
but one of the 16 QSOs have both G130M and G160M spectra available.
Although only G160M spectra are available in the {\it HST} archive for
SDSSJ\,124410.82$+$172104.6, the LRG in this field occurs at
sufficiently high redshift ($z_{\rm LRG}=0.5591$) that the G160M alone
already provides the full coverage for its associated Lyman series
transitions.

For each QSO, individual one-dimensional spectra produced by the {\it
  CALCOS} pipeline were retrieved from the {\it HST} archive, and
further processed using our own software.  The wavelength calibration
for COS spectra has been shown to contain wavelength-dependent errors
up to $\approx 15$ \kms\ (e.g., Johnson \etal\ 2013; Liang \& Chen
2014; Wakker \etal\ 2015), leading to apparent misalignments of
spectral features between different exposures.  Additional wavelength
calibration effort is therefore needed both to optimize the
signal-to-noise ($S/N$) in the final combined spectrum and to ensure
the accuracy in measuring the kinematic properties of different
absorption features.

We developed our own software to perform this necessary wavelength
calibration in two steps.  First, we corrected the {\it relative}
wavelength offsets between individual spectra using a low-order
polynomial that best describes the offsets between common absorption
lines found in different exposures.  The common absorption lines
include both intervening absorbers at $z>0$ and strong Milky Way
features, such as \SiIII\,$\lambda\,1206$ and \CII\,$\lambda\,1334$ in
the G130M data and \SiII\,$\lambda\,1526$ and \AlII\,$\lambda\,1670$
in the G160M data.  Individual wavelength-corrected spectra were then
coadded to form a final combined spectrum.  Next, an {\it absolute}
wavelength correction was performed using a low-order polynomial by
registering unsaturated low-ionization Milky Way features to their
known wavelengths in vacuum.  In general, the absolute wavelength
calibration for the G160M spectra was more challenging than the G130M
data due to the fact that fewer unsaturated lines were available.  The
final solution was evaluated and verified using available Lyman series
of strong intervening \HI\ absorbers across the full spectral range.
In some cases, a new solution for the G160M spectrum was obtained
using only a constant wavelength offset that was determined based on
absorption features in the overlapping window between the G130M and
G160M data.  When comparing the velocity centroids between
low-ionization lines observed in COS and those of
\MgII\,$\lambda\lambda\,2796, 2803$ lines observed in ground-based
echelle spectra (${\rm FWHM}\approx 10$ \kms; see the discussion in
\S\ 2.3), we found that the final wavelength solution generated from
our software is accurate to within $\pm 3$ \kms.  Finally, each
combined spectrum was continuum normalized using a low-order
polynomial fit to spectral regions free of strong absorption features.
The final continuum-normalized spectra have a median $S/N \approx
10-30$ per resolution element.

\subsection{Optical Echelle Spectroscopy}

Optical echelle spectra of the background QSOs are available for 11 of
the 16 QSOs in the COS-LRG sample.  These echelle spectra extend the
wavelength coverage of absorption spectroscopy from $\lambda\apg 3300$
\AA\ to $\lambda\apg 5000$ \AA, providing additional constraints based
on observations of the \FeII\ absorption series, the
\MgII\,$\lambda\lambda\,2796, 2803$ doublet features, and
\MgI\,$\lambda\,2852$ absorption.  Specifically, optical echelle
spectra of SDSSJ\,024651.91$-$005930.9, SDSSJ\,112756.76$+$115427.1,
SDSSJ\,124410.82$+$172104.6, SDSSJ\,135726.27$+$043541.4, and
SDSSJ\,140626.60$+$250921.0 were obtained using the MIKE echelle
spectrograph (Bernstein \etal\ 2003) on the Magellan Clay telescope.
The observations were carried out in three separate runs in September
2016, January 2017, and April 2017.  We used a $1''$ slit and $2\times
2$ binning during readout, resulting in a spectral resolution of ${\rm
  FWHM}\approx 11$ \kms at wavelength $\lambda<5000$ \AA.  The echelle
data were processed and extracted using a custom software described in
Chen \etal\ (2014) and in Zahedy \etal\ (2016).  Wavelengths were
calibrated using a ThAr frame obtained immediately after each science
exposure and subsequently corrected to vacuum and heliocentric
wavelengths.  Relative flux calibrations were performed using a
sensitivity function determined from a spectrophotometric standard
star observed on the same night as the QSOs.  Individual
flux-calibrated echelle orders from different exposures were then
coadded and combined to form a single final spectrum that covers a
wavelength range from $\lambda\approx 3300$ \AA\ to $\lambda\approx
9500$ \AA\ for each QSO.  Finally, the combined spectrum was continuum
normalized using a low-order polynomial fit to the spectral regions
free of strong absorption features.

Optical echelle spectra of SDSSJ\,080359.23$+$433258.3,
SDSSJ\,091029.75$+$101413.5, SDSSJ\,092554.70$+$400414.1,
SDSSJ\,095000.73$+$483129.2, SDSSJ\,155048.29$+$400144.9, and
SDSSJ\,155304.92$+$354828.6 obtained using HIRES (Vogt \etal\ 1994) on
the Keck Telescopes are available in the Keck Observatory Archive
(KOA).  The spectra were obtained using a $0.86''$ slit and $2\times
1$ binning during readout, resulting in a spectral resolution of ${\rm
  FWHM}\approx 6.3$ \kms at wavelength $\lambda<5900$ \AA.  Individual
pipeline-reduced echelle orders were retrieved from the KOA, and
coadded using our own software to form a single combined echelle
spectrum, covering a wavelength range from $\lambda\approx 3100$
\AA\ to $\lambda\approx 5900$ \AA\ for each QSO.  The combined
spectrum was continuum normalized using a low-order polynomial fit to
the spectral regions free of strong absorption features.  A summary of
available optical echelle spectra is presented in Table 3, which lists
for each QSO the echelle spectrograph used, the total accumulated
exposure time, and the mean $S/N$ per resolution element at $\lambda <
5000$ \AA.

\begin{table}
\scriptsize
\centering
\caption{Journal of Optical Echelle Observations}
\label{table:mikeobs}
\centering {
\begin{tabular}{lrrr}
\hline \hline
\multicolumn{1}{c}{QSO} & \multicolumn{1}{c}{Instrument} & \multicolumn{1}{c}{$t_{\rm exp}$ (s)} & \multicolumn{1}{c}{$S/N$} \\
\hline 
SDSSJ\,024651.91$-$005930.9 &  MIKE & 3000 & 41 \\
SDSSJ\,080359.23$+$433258.3 & HIRES & 6000 & 24 \\
SDSSJ\,091029.75$+$101413.5 & HIRES & 2400 & 15 \\
SDSSJ\,092554.70$+$400414.1 & HIRES & 2400 & 16 \\
SDSSJ\,095000.73$+$483129.2 & HIRES & 2400 & 30 \\
SDSSJ\,112756.76$+$115427.1 &  MIKE & 7620 & 17 \\
SDSSJ\,124410.82$+$172104.6 &  MIKE & 4200 & 33 \\
SDSSJ\,135726.27$+$043541.4 &  MIKE & 5400 & 25 \\
SDSSJ\,140626.60$+$250921.0 &  MIKE & 1800 & 10 \\
SDSSJ\,155048.29$+$400144.9 & HIRES & 2400 & 31 \\
SDSSJ\,155304.92$+$354828.6 & HIRES & 2400 & 36 \\
\hline
%		\hline \multicolumn{12}{l}{\bf Notes} \\
%\multicolumn{11}{l}{$^\mathrm{a}$COS-Halos red galaxies that match our LRG selection criteria.} \\
\end{tabular}
}
\end{table}

\section{Absorption Properties in LRG Halos}

The continuum-normalized UV and optical QSO spectra described in \S\ 2
provide sensitive constraints for the properties of absorbing gas in
massive quiescent halos.  For each LRG halo, we search for
corresponding absorption features in the spectrum of the background
QSO over a line-of-sight velocity interval of $\pm 500$ \kms\ from the
systemic redshift of the LRG.  This search window corresponds to
$\approx \pm\,3\,\sigma_v$, where $\sigma_v$ is the observed
line-of-sight velocity dispersion of \MgII\ absorbing gas in LRG halos
(e.g., Huang \etal\ 2016).  It is therefore sufficiently large to
include absorption features physically connected to the LRGs. 

To characterize the gas content in LRG halos, we obtain two sets of
measurements.  First, we perform a Voigt profile analysis, taking into
account the presence of multiple absorption components per halo, and
determine $N(\HI)$ and the Doppler parameter $b_{\HI}$ of individual
components, and the associated measurement uncertainties.  Next, we
measure the total, integrated absorption equivalent widths of
different ionic absorption transitions, which enable direct
comparisons between absorption properties of the LRG halos in our
study and those of quiescent halos studied previously (e.g.\ Gauthier
\& Chen 2011; Thom \etal\ 2012; Werk \etal\ 2013; Huang \etal\ 2016).
Here we describe the absorption line analysis for measuring $N(\HI)$
and for constraining the incidence of heavy ions in massive quiescent
halos.

\begin{subfigures}
\begin{figure*}
\begin{adjustbox}{
addcode=
{\begin{minipage}{\width}}
{\caption{The Lyman series absorption spectra observed at different
  projected distances from LRGs, from $d=42$ kpc at the top and
  increasing to $d=102$ kpc at the bottom.  The continuum-normalized
  spectra are shown in black with the corresponding 1-$\sigma$ error
  shown in cyan.  The green and red dash-dotted lines mark the
  normalized continuum and zero flux levels for guidance.  For each
  LRG halo, the velocity profiles of Ly$\gamma$, and Ly$\delta$ are
  presented in the two right panels with zero velocity corresponding
  to $z_{\rm abs}$ in Table 4, and the remaining higher-order Lyman
  series lines, along with the Lyman limit, are presented in the left
  panel with the blue dotted lines indicating the expected positions
  of the Lyman transitions.  Spectral regions that are not covered by
  available COS spectra are greyed out.  The best-fit Lyman series
  absorption spectra are shown in red} % with
%  individual components displayed in different colors.}
        \label{figure:LLS}\end{minipage}},
rotate=90,center}
\includegraphics[scale=0.65]{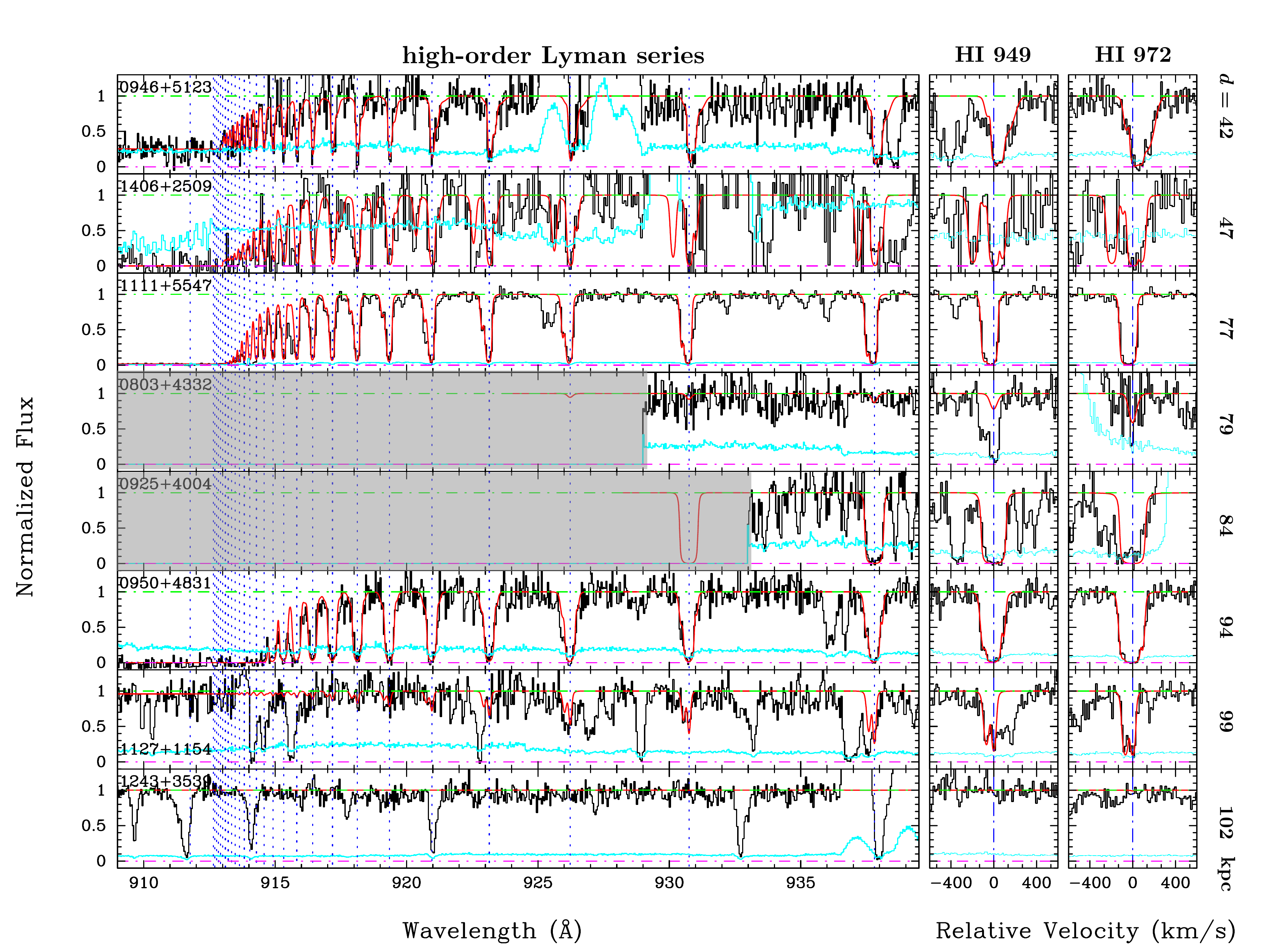}
\end{adjustbox}
\end{figure*}
\begin{figure*}
\begin{adjustbox}{
addcode=
{\begin{minipage}{\width}}
{\caption{Same as Figure 2a, but for LRG halos at projected distances
  from $d=107$ kpc at the top and increasing to $d=160$ kpc at the
  bottom.  Note that for SDSSJ\,155048.29$+$400144.9 in the top panel
  the best-fit $N(\HI)$ successfully reproduces the full Lyman series
  lines but predicts a larger flux decrement below the Lyman limit
  than what is observed in the data by more than 1-$\sigma$.  The
  acquisition image of the COS observations reveals a second point
  source in the COS aperture at $\approx 0.55''$ along the dispersion
  direction from the QSO, which may have contributed to the excess
  flux below the Lyman limit of the LRG.}
        \label{figure:LLS}\end{minipage}},
rotate=90,center}
\includegraphics[scale=0.65]{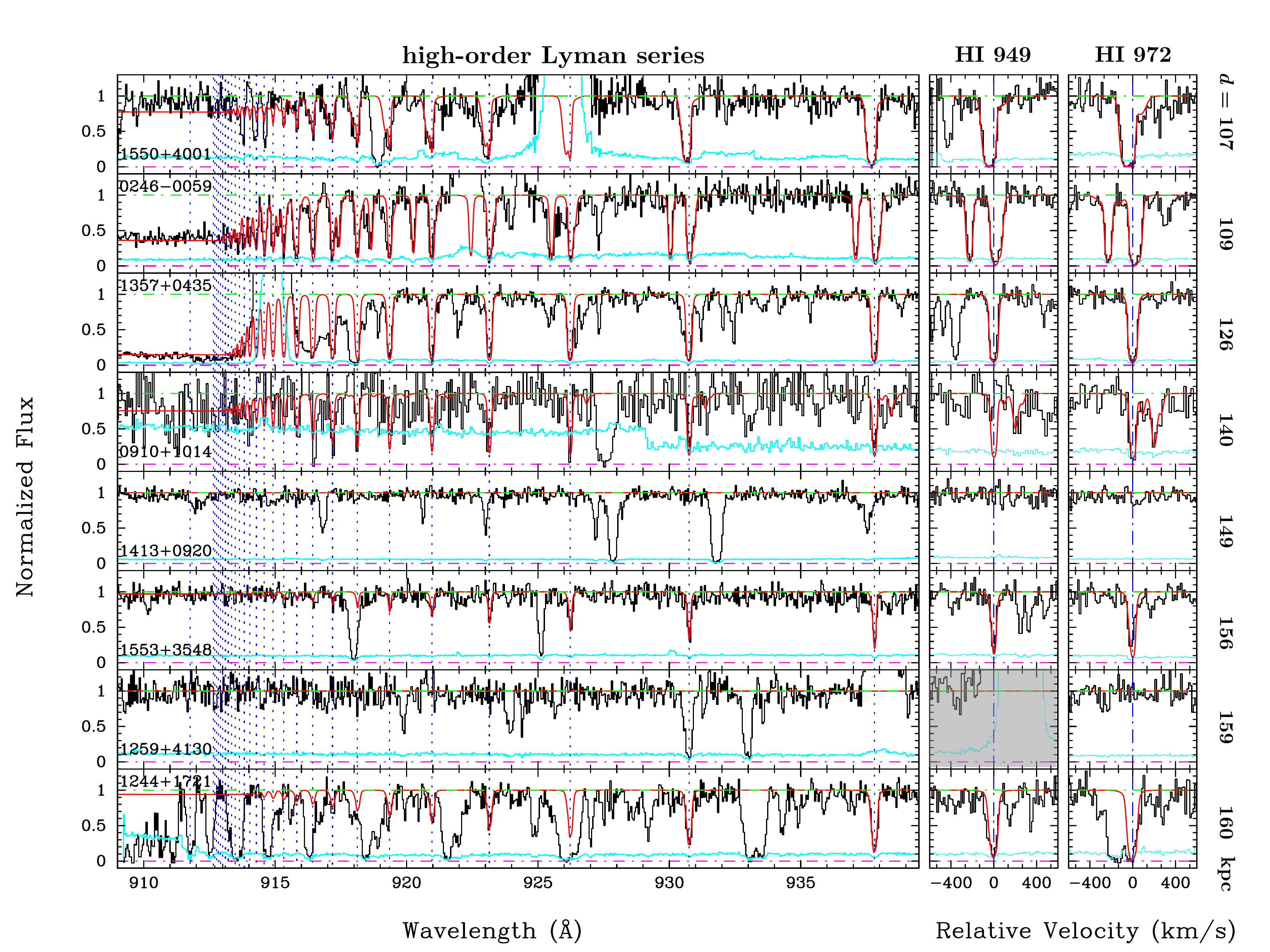}
\end{adjustbox}
\end{figure*}
\end{subfigures}

\subsection{Measurements of $N(\HI)$ and $b_{\scriptsize {\rm HI}}$}

The COS-LRG sample is established based on the criterion that robust
constraints for $N({\HI})$ can be placed based on observations of the
Lyman series lines.  Specifically, observations of the flux decrement
at the Lyman limit transition provide a direct constraint for
$N(\HI)$, while observations of the Lyman absorption lines constrain
the Doppler parameter $b_{\HI}$.  In Figures 2a\&b, we present the
continuum-normalized final spectra of the LRG halos over the
rest-frame spectral range from $\lambda\apl 910$ \AA\ to Ly$\gamma$,
organized with increasing projected distance from top to bottom.

For all but two sightlines (SDSSJ\,080359.23$+$433258.3 and
SDSSJ\,092554.70$+$400414.1), the available COS data provide a
complete spectral coverage for the Lyman series transitions from
\lya\ to below Lyman limit\footnote{The LRG toward
  SDSSJ\,124409.17$+$172111.9 is found at $z=0.5591$ and the
  corresponding \lya\ transition occurs at $\lambda=1895$ \AA.
  Therefore, only \lyb\ and higher Lyman series lines are observed in
  the COS FUV channel.}.  For SDSSJ\,080359.23$+$433258.3, the LRG is
at $z=0.2535$ and available COS spectra do not provide spectral
coverage below $\lambda\approx 930$ \AA\ at the rest frame of the LRG.
However, because the absorbing gas along the QSO sightline in this LRG
halo is optically thin, observations of \lya, \lyb, Ly$\gamma$ and
Ly$\delta$ are sufficient for constraining the $N({\HI})$ without
additional higher-order Lyman series lines.  For
SDSSJ\,092554.70$+$400414.1, the LRG occurs at $z=0.2475$ and
therefore available COS spectra also do not cover the Lyman limit and
higher-order Lyman series lines.  However, because the cool gas
uncovered along the QSO sightline contains a large neutral gas column
and the \lya\ absorption line exhibits prominent damping wings, an
accurate measurement of $N(\HI)$ can also be obtained based on the
damped profiles.

For each LRG halo, we perform a Voigt profile analysis using a custom
software.  We first generate a model absorption spectrum, including
the full Lyman series and continuum absorption beyond the Lyman limit,
based on the minimum number of discrete components needed to explain
the observed absorption profiles.  Each component is uniquely defined
by three parameters: $N(\HI)$, $b_{\HI}$, and the velocity centroid
$dv_c$ relative to the redshift of the strongest absorbing component
$z_{\rm abs}$.  The theoretical Lyman series spectrum is then
convolved with the COS line spread function, appropriate for the
lifetime position the spectra were obtained at, and binned to the
adopted pixel resolution of the data.  Next, the model spectrum is
compared to the continuum-normalized COS spectrum, and the best-fit
model is found using a $\chi^2$ minimization routine.  Finally, we
perform a Markov Chain Monte Carlo (MCMC) analysis using the
\textsc{emcee} package (Foreman-Mackey \etal\ 2013), in order to
assess the probability distribution of the best-fit model parameters.
The MCMC run consists of 200 steps with an ensemble of 250 walkers,
which are initialized over a small region in the parameter space
around the minimum $\chi^2$ value.  The MCMC analysis allows us to
construct the marginalized probability distribution for each parameter
of the model.  The sum of the best-fit Voigt profiles of individual
components is displayed in red solid line for each LRG halo in Figures
2a\&b.  The best-fit $N(\HI)$ and $b_{\HI}$ of each component and the
corresponding 68\% confidence intervals are presented in columns (8)
and (9) of Table 4, along with $z_{\rm abs}$, relative velocity
$v_{{\rm gas}-{\rm gal}}$ of $z_{\rm abs}$ to $z_{\rm LRG}$, $dv_c$,
and the total $N(\HI)$ summed over all components in columns (4)
through (8).  Note that while the search window in the line-of-sight
velocity is large, $\pm 500$ \kms, all the identified absorption
features fall well within this search window with a mean and
dispersion between the strongest \HI\ component and the LRG of
$\langle\,v_{{\rm gas}-{\rm gal}}\rangle=29$ \kms\ and
$\sigma_{\langle\,v_{{\rm gas}-{\rm gal}}\rangle}=171$ \kms,
consistent with the velocity dispersion found between \MgII\ absorbing
gas and LRGs in the larger SDSS LRG sample in Zhu \etal\ (2014) and in
Huang \etal\ (2016).

Three of the 16 LRGs do not exhibit any corresponding Lyman absorption
series in the search window.  These are SDSSJ\,124307.36$+$353926.3 at
$z=0.3896$ and $d=102$ kpc, SDSSJ\,141307.39$+$091956.7 at $z=0.3584$
and $d=149$ kpc, and SDSSJ\,125859.98$+$413128.2 at $z=0.2790$ and
$d=159$ kpc.  We place a sensitive 2-$\sigma$ upper limit of
$N(\HI)<12.8$ over a wavelength window of $\pm 16$ \kms\ around the
expected \lya\ location, approximately twice the spectral resolution
element of the COS spectra, using the associated 1-$\sigma$ error
spectrum for these LRG halos.

Our analysis shows that high-$N(\HI)$ gas is common in LRG halos with
a median of $\langle\,\log\,N(\HI)\,\rangle_{\rm med}=16.6$ at $d\apl
160$ kpc.  The spectral coverage of the rest-frame Lyman limit
transition proves to be critical for constraining $N(\HI)$ in these
Lyman limit and partial Lyman limit systems.  The excellent agreement
between the observed and best-fit absorption profiles of the Lyman
series lines and the flux discontinuity at the Lyman limit in Figure
2a\&b demonstrates that our best estimated $N(\HI)$ and $b_{\HI}$ in
Table 4 are indeed accurate.  In one case, SDSSJ\,155048.29$+$400144.9
at $z=0.3125$ and $d=107$ kpc, we note that the best-fit $N(\HI)$
successfully reproduces the full Lyman series lines but predicts a
larger flux decrement below the Lyman limit than what is observed in
the data by more than 1-$\sigma$.  Inspecting the acquisition image of
the COS observations, we find a second point source in the COS
aperture at $\approx 0.55''$ along the dispersion direction from the
QSO.  The excess flux below the Lyman limit of the LRG can be
explained if the redshift of this second source is below the redshift
of the LRG.  We also note a second case, SDSSJ\,124409.17$+$172111.9
at $z=0.5591$ and $d=160$ kpc, for which the continuum flux declines
to zero below 909 \AA\ in the rest frame of the LRG.  A
\MgII\ absorber of rest-frame equivalent width $\ewmgii\approx 0.4$
\AA\ at $z=0.5507$ toward this QSO sightline has been known and
studied extensively by several groups (e.g., Steidel \& Sargent 1992;
Rao \etal\ 1995; Churchill \etal\ 2001; Zahedy \etal\ 2017b).  This
\MgII\ absorber contains a large neutral gas column of
$\log\,N(\HI)=19.1\pm 0.1$ as revealed in the COS spectra, which
explains the flux decrement below 909 \AA\ in the rest frame of the
$z=0.5591$ absorber.

\subsection{Absorption Equivalent Width of Heavy Ions}

In addition to the hydrogen Lyman series absorption features, we also
search for associated ionic transitions for constraining the incidence
of heavy ions in massive quiescent halos, including
$\MgII\,\lambda\lambda\,2796, 2803$, $\SiII\,\lambda\,\,1260$,
$\SiII\,\lambda\,1193$, $\SiIII\,\lambda\,1206$,
$\CIII\,\lambda\,977$, and $\OVI\,\lambda\lambda\,1031, 1037$.
Figures 3a\&b presents the absorption profiles of associated
\MgII\ doublet, \SiII, \SiIII, \CIII, and the \OVI\ doublet
transitions found in each LRG halo, along with the hydrogen \lya\ and
\lyb\ absorption lines, both observations and best-fit Voigt profile
models, at the top of each column for comparison.  For the three LRGs
with no \HI\ absorption detected, we included the Voigt profiles for
absorbing gas of $\log\,N(\HI)=12.8$ and $b_{\HI}=15$ \kms\ for
comparison.  Zero velocity corresponds to $z_{\rm abs}$ in Table 4.
For clarity, spectral regions that are not covered by available COS
spectra or contain contaminating features are greyed out.  For \HI,
\MgII, \SiII, and \OVI\ lines, two transitions are observed which
enable identifications of contaminating features based on their known
relative line strengths.  For \SiIII\ and \CIII, we rely on the
matching velocity profiles with other low-ionization lines to identify
contaminating features.

We note the excellent agreement in the velocity centroids between
individual components of different ionic features, such as the
\MgII\ absorption doublets and \SiII, \SiIII, and \CIII.  Recall the
issues with COS wavelength calibration described in \S\ 2.2, which
were corrected using known Milky Way absorption features in our custom
data reduction pipeline by forcing the Milky Way lines to their known
wavelengths in vacuum.  Consequently, an implicit assumption of this
process is zero relative motion between the local standard of rest
(LSR) and the ISM gas that dominates these absorption features.  Such
assumption does not exist in ground-based echelle spectra of
\MgII\ absorption lines, which were calibrated using a combination of
comparison lamp and sky lines.  The observed excellent agreement
between \MgII\ absorption lines and COS observations of \SiII, \SiIII,
and \CIII\ lines therefore indicates that the relative motion between
the Milky Way gas and the LSR toward these sightlines is indeed
negligible.  In addition, the matching velocity profiles between \HI,
\MgII, and the remaining FUV lines also provide an important
additional guide for filtering out contaminating features.

Two features are immediately clear from Figures 3a\&b.  First of all,
heavy ions are common in these massive quiescent halos.  In all but
one LRG with \lya\ and \lyb\ detected, metal absorption features are
also detected (see \S\ 4.2 for more discussion).  Secondly, while the
velocity spread of all ionic absorption components are seen to extend
up to $\Delta\,v\approx 200$ \kms\ in three cases, e.g.,
SDSSJ\,094632.40$+$512335.9 at $z=0.4076$ and $d=42$ kpc,
SDSSJ\,024651.20$-$005914.1 at $z=0.4105$ and $d=109$ kpc, and
SDSSJ\,091029.75$+$101413.5 at $z=0.2641$ and $d=140$ kpc, the line
profiles of individual components for all LRG halos are narrow.  The
observed narrow component profiles suggest that the gas is
photo-ionized.  The only exception is the \OVI\ doublet transitions
detected in the halo of SDSSJ\,094632.40$+$512335.9, which appear
broad (${\rm FWHM}\approx 300$ \kms) and asymmetric, indicating a
different physical origin from the associated hydrogen and
low-ionization gas.

We measure the rest-frame total, integrated absorption equivalent
widths of both \lya\ and associated ionic lines, including
$\MgII\,\lambda\,2796$, $\SiII\,\lambda\,\,1260$,
$\SiIII\,\lambda\,1206$, $\CIII\,\lambda\,977$, and
$\OVI\,\lambda\,1031$, that are not contaminated by other absorption
systems.  A local continuum is determined using the line-free spectral
region outside of the designated line features, in order to minimize
systematic uncertainties due to continuum placement error.  Equivalent
width measurement uncertainties are then determined based on the
associated 1-sigma error spectrum.  For non-detections, we place a
2-$\sigma$ upper limit to the underlying absorption feature over a
wavelength window of $\pm 16$ \kms\ around the expected line location,
using the associated error spectrum.  The wavelength window is chosen
to cover approximately twice the spectral resolution element of the
COS spectra around the expected lines.  The absorption equivalent
width measurements and associated uncertainties are presented in
columns (9) through (14) of Table 4, along with 2-$\sigma$ upper
limits for non-detections.  Entries with ``...''  indicate that either
the absorption transitions are not covered by available data or they
are blended with contaminating features and therefore no equivalent
width measurements are possible.  For the five LRGs that have also
been studied by the COS-Halos team, equivalent width measurements of
these transitions have also been published in Werk \etal\ (2013).
Good agreements, to within 2-$\sigma$ measurement uncertainties are
found in all but one transition.  For SDSSJ\,091029.75$+$101413.5 at
$z=0.2641$ and $d=140$ kpc, Werk \etal\ (2013) reported $W_r=282\pm
63$ m\AA\ for $\CIII\,\lambda\,977$, but we measure $W_r(977)=681\pm
48$ m\AA.  Comparing the continuum-normalized absorption profiles of
our own and those presented in Figure 3 of Werk \etal\ (2013), we find
that the discrepancy is likely due to a combination of low $S/N$ and
uncertainties in the continuum normalization.

\section{Discussion}
\label{section:sample}

The absorption-line search described in \S\ 3 has uncovered strong
\HI\ absorbers in a large fraction of LRG halos at $z=0.21-0.55$.
Observations of the full hydrogen Lyman series have enabled accurate
and precise measurements of $N(\HI)$.  The median \HI\ column density
is found to be $\langle\,\log\,N(\HI)\rangle = 16.6$ at $d\apl 160$
kpc (or $d \apl 1/3\,R_{\rm vir}$) in these LRG halos.  The
line-of-sight velocity separations between the \HI\ absorbing gas and
LRGs are characterized by a mean and dispersion of $\langle\,v_{{\rm
    gas}-{\rm gal}}\rangle=29$ \kms\ and $\sigma_{\langle\,v_{{\rm
      gas}-{\rm gal}}\rangle}=171$ \kms.  The observed velocity
dispersion of \HI\ gas is similar to what is observed for
\MgII\ absorbing gas (e.g., Zhu \etal\ 2014; Huang \etal\ 2016), but
less than the expected line-of-sight velocity dispersion for
virialized gas in these massive haloes.  A suppressed velocity
dispersion implies that the kinetic energy of the absorbing clumps is
being dissipated, possibly due to the ram-pressure drag force as the
clumps move through the hot halo.  We refer the readers to Gauthier \&
Chen (2011) and Huang \etal\ (2016) for a discussion on the implied
mass limit for the clumps.

In addition to accurate $N(\HI)$ measurements, we have also been able
to obtain accurate constraints for $b_\HI$ of individual components.
The best-estimated $b_\HI$ values for optically-thick components
($\log\,N(\HI)>17.2$) are all relatively narrow with $b_\HI \apl 20$
\kms, indicating a relatively cool gas temperature of $T\apl 3\times
10^4$ K.  Combining the observed \HI\ absorption profiles with those
of low-, intermediate-, and high-ionization species provides a unique
opportunity for constraining the ionization condition, thermal state,
and chemical abundances of the CGM in massive quiescent halos.  A
detailed ionization analysis is presented in a subsequent paper
(Zahedy \etal\ 2018, in preparation).  Here we discuss general halo
gas properties observed around LRGs.

\begin{subfigures}
\begin{figure*}
\begin{adjustbox}{
addcode=
{\begin{minipage}{\width}}
{\caption{Absorption profiles of (from top to bottom) \lya, \lyb, and
  associated \MgII, \SiII, \SiIII, \CIII, and \OVI\ for eight LRG
  halos at $d<105$ kpc (increasing $d$ from left to right).  Following
  Figure 2, zero velocity corresponding to $z_{\rm abs}$ in Table 4,
  1-$\sigma$ errors are shown in cyan.  Spectral regions that are
  either contaminated or not covered by available absorption spectra
  are greyed out for clarity.  Systemic velocities of the LRGs are
  marked by the vertical dotted line, and the spectral windows used
  for equivalent width measurements in Table 4 are marked with a
  horizontal bar at the top of each panel.  For the LRG at $d=102$
  kpc, no absorption features are detected and 2-$\sigma$ upper limits
  are obtained within $\pm 1$ spectral resolution element, the window
  marked by the thick horizontal bar.  For the broad damped
  \lya\ feature at $d=84$ kpc, $W_r(1215)$ was measured from $-800$ to
  600\,\kms.  The best-fit \lya\ and \lyb\ Voigt profiles, both for
  individual components separately (thin lines) and for all components
  combined (thick red line), of each LRG halo are also reproduced in
  the top two panels for direct comparisons with resolved metal-line
  components.  All LRGs at $d<100$ kpc exhibit both hydrogen and
  associated metal-line absorption features.}
%, suggesting a modest level
%  of chemical enrichment in these LRG halos. }
\label{figure:ions}\end{minipage}},
rotate=90,center}
\includegraphics[scale=0.6]{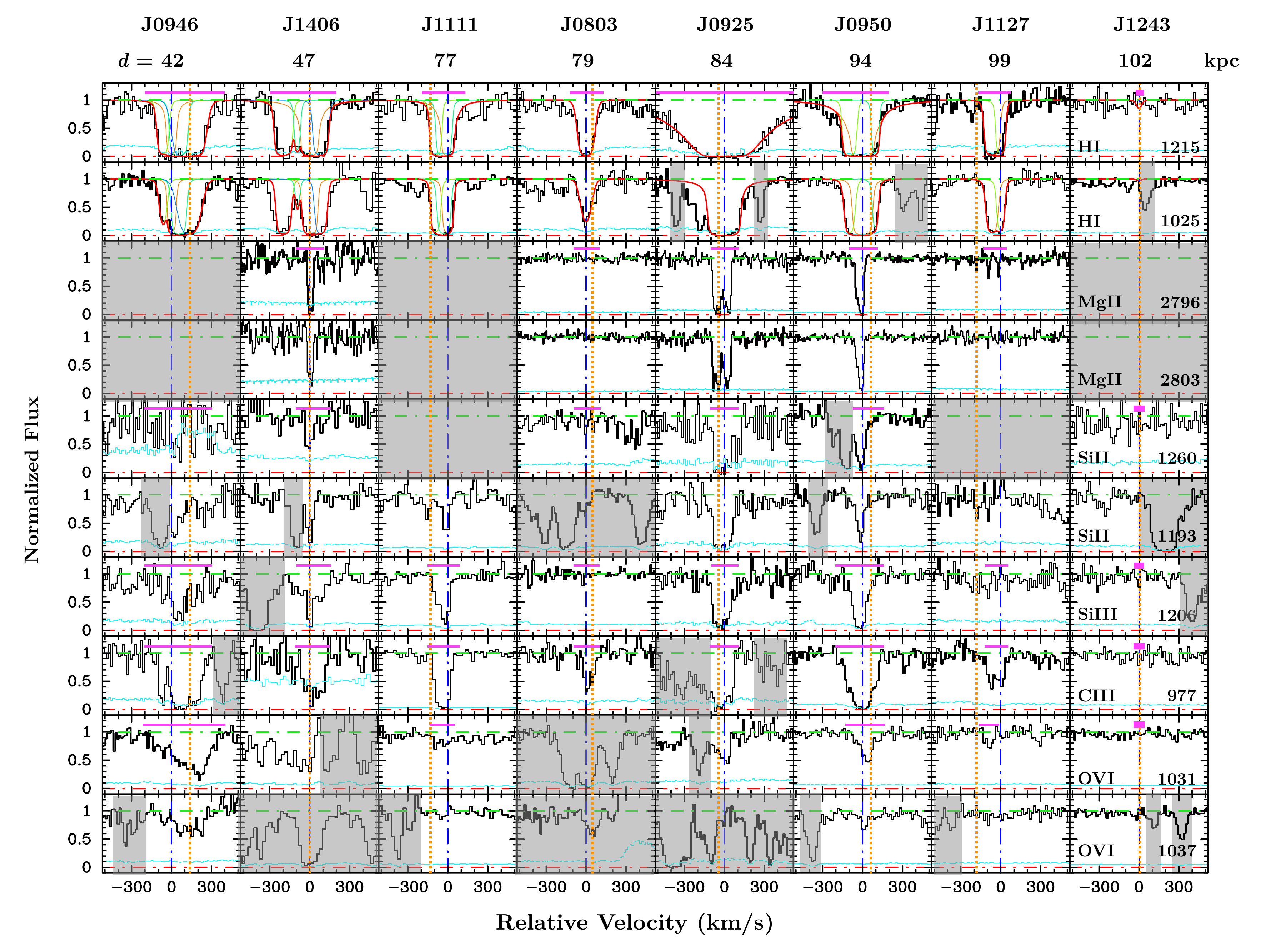}%COS_LRG_vprof_components.pdf}
\end{adjustbox}
\end{figure*}
\begin{figure*}
\begin{adjustbox}{
addcode=
{\begin{minipage}{\width}}
{\caption{Same as Figure 3a, but for the remaining eight LRG halos at
  $d=107-160$ kpc.  For the LRGs at $d=149$ and 159 kpc, no absorption
  features are detected and 2-$\sigma$ upper limits are obtained
  within $\pm 1$ spectral resolution element, the window marked by the
  thick horizontal bar.  The LRG at $d=126$ kpc is the only one of all
  13 \lya-absorbing LRGs in our sample with no detectable metal-line
  absorption features.}
\label{figure:ions}\end{minipage}},
rotate=90,center}
\includegraphics[scale=0.65]{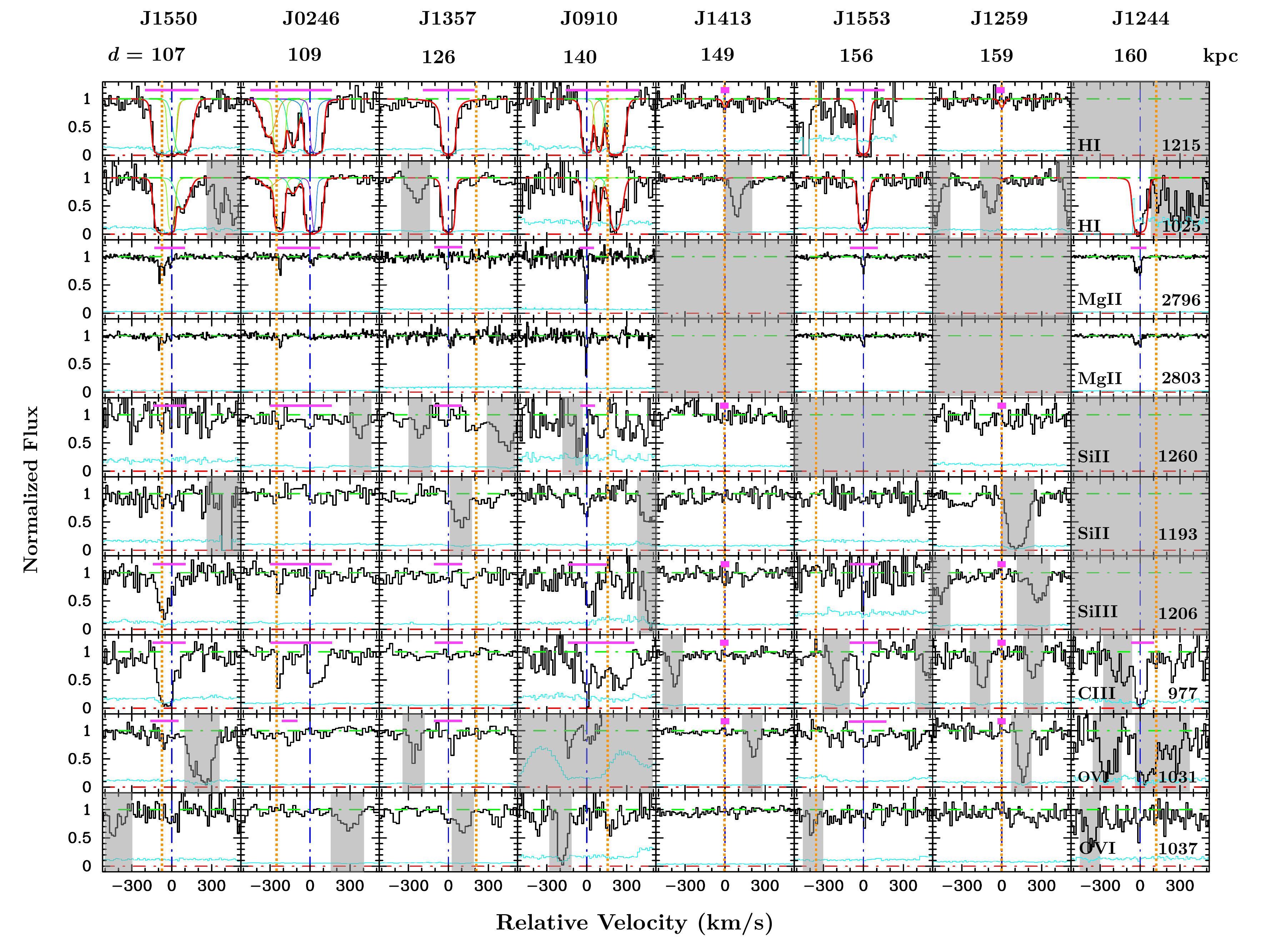}%COS_LRG_vprof_components.pdf}
\end{adjustbox}
\end{figure*}
\end{subfigures}

\begin{landscape}
\begin{table}
\centering
\caption{Summary of LRG Halo Gas Properties}
\label{table:measurements}
\resizebox{9in}{!}{
%\centering {
\begin{tabular}{lrrcrrrrllrrrrrr}
\hline \hline
\multicolumn{3}{c}{LRG} & & \multicolumn{12}{c}{Absorption Properties} \\
\cline{1-3} 
\cline{5-16}
                        &                                   & \multicolumn{1}{c}{$d$}  & &  & \multicolumn{1}{c}{$v_{{\rm gas}-{\rm gal}}$} &                                 & \multicolumn{1}{c}{$dv_c$} &  & \multicolumn{1}{c}{$b_{\HI}$} & \multicolumn{1}{c}{\ewlya} & \multicolumn{1}{c}{\ewciii} & \multicolumn{1}{c}{\ewovi}  & \multicolumn{1}{c}{\ewsiiii}  & \multicolumn{1}{c}{\ewsiii} & \multicolumn{1}{c}{\ewmgii} \\
\multicolumn{1}{c}{ID} &  \multicolumn{1}{c}{$z_{\rm LRG}$} & \multicolumn{1}{c}{(kpc)} & &\multicolumn{1}{c}{$z_{\rm abs}$}& \multicolumn{1}{c}{(\kms)} & \multicolumn{1}{c}{comp.} & \multicolumn{1}{c}{(\kms)} & \multicolumn{1}{c}{\lnhone} & \multicolumn{1}{c}{(\kms)} & \multicolumn{1}{c}{(m\AA)} & \multicolumn{1}{c}{(m\AA)} & \multicolumn{1}{c}{(m\AA)} & \multicolumn{1}{c}{(m\AA)} & \multicolumn{1}{c}{(m\AA)} & \multicolumn{1}{c}{(m\AA)}  \\
\multicolumn{1}{c}{(1)} & \multicolumn{1}{c}{(2)} & \multicolumn{1}{c}{(3)} & & \multicolumn{1}{c}{(4)} & \multicolumn{1}{c}{(5)} & \multicolumn{1}{c}{(6)} & \multicolumn{1}{c}{(7)} & \multicolumn{1}{c}{(8)} & \multicolumn{1}{c}{(9)} & \multicolumn{1}{c}{(10)} & \multicolumn{1}{c}{(11)} & \multicolumn{1}{c}{(12)} & \multicolumn{1}{c}{(13)} & \multicolumn{1}{c}{(14)} & \multicolumn{1}{c}{(15)} \\
\hline 
SDSSJ\,094632.40$+$512335.9 & 0.4076 &  41.7 & & 0.40701 & $-126$ & all &  ... &$17.34\pm 0.01$  & ... & $1639\pm 47$ & $926\pm 37$ & $800\pm 27$ & $688\pm 54$ &    $<\,420$  &  ... \\
 &  &   & &&&1&  $-63.8^{+4.5}_{-4.1}$ & $14.69^{+0.09}_{-0.08}$ & $27.5^{+3.8}_{-3.6}$  & ... & ... & ... & ... &    ...  &  ... \\
 &  &   & &&&2&  $0.0^{+0.4}_{-0.3}$ & $17.32\pm 0.01$ & $12.3^{+0.8}_{-0.7}$& ... & ... & ... & ... &    ...  &  ... \\
 &  &   & &&&3&  $+54.6^{+4.0}_{-5.5}$ & $16.01^{+0.09}_{-0.10}$ & $29.9^{+3.7}_{-3.4}$ & ... & ... & ... & ... &    ...  &  ... \\
 &  &   & &&&4&  $+132.1^{+10.7}_{-18.6}$ & $15.41^{+0.14}_{-0.09}$ & $71.4^{+9.8}_{-6.0}$ & ... & ... & ... & ... &    ...  &  ... \\

SDSSJ\,140625.97$+$250923.2 & 0.4004 &  47.3 & & 0.40040 & $0$ & all&   ... & $18.04^{+0.14}_{-0.04}$ &... & $1535\pm 40$ & $610\pm 140$ &     ...     & $433\pm 30$ & $192\pm 39$ & $431\pm 67$ \\
 &  &   & &&&1&  $-196.9^{+3.2}_{-3.0}$ & $16.15\pm0.22$ & $28.6^{+2.4}_{-1.9}$& ... & ... & ... & ... &    ...  &  ... \\
 &  &   & &&&2&  $-93.3^{+5.6}_{-4.8}$ & $14.15^{+0.14}_{-0.17}$ & $16.6^{+2.6}_{-2.2}$ & ... & ... & ... & ... &    ...  &  ... \\
 &  &   & &&&3&  $0.0^{+0.1}_{-0.1}$ & $18.03^{+0.15}_{-0.04}$ & $20.2^{+1.0}_{-1.2}$ & ... & ... & ... & ... &    ...  &  ... \\
 &  &   & &&&4&  $+82.7^{+4.0}_{-5.4}$ & $15.68^{+0.23}_{-0.22}$ & $26.8^{+3.0}_{-2.4}$ & ... & ... & ... & ... &    ...  &  ... \\

SDSSJ\,111132.33$+$554712.8 & 0.4629 &  77.1 & & 0.46352 & $+127$ & all& ... & $17.82\pm0.01$ & ... &  $876\pm 27$ & $401\pm 5$  &  $67\pm 10$ & $362\pm 21$ &     ...     & ... \\
 &  &   & &&&1&  $-92.1^{+1.6}_{-1.3}$ & $15.73\pm0.03$ & $20.1^{+1.1}_{-1.0}$& ... & ... & ... & ... &    ...  &  ... \\
 &  &   & &&&2&  $-30.4^{+2.5}_{-3.0}$ & $16.32^{+0.06}_{-0.08}$ & $23.9^{+1.7}_{-2.0}$ & ... & ... & ... & ... &    ...  &  ... \\
 &  &   & &&&3&  $0.0^{+0.6}_{-0.8}$ & $17.80\pm0.01$ & $14.8^{+0.4}_{-0.3}$ & ... & ... & ... & ... &    ...  &  ... \\

SDSSJ\,080357.74$+$433309.9 & 0.2535 &  78.5 & & 0.25330 & $-48$ & 1 &$0.0^{+2.4}_{-3.0}$ & $14.77\pm 0.05$ & $42.7^{+3.8}_{-2.4}$ &  $591\pm 18$ & $175\pm 23$ &     ...     &     $<\,48$ &     $<\,68$ & $<\,29$ \\

SDSSJ\,092554.18$+$400353.4 & 0.2475 &  83.7 & & 0.24769 & $+46$ & 1 &  $0.0\pm1.7$ & $19.58\pm 0.02$ & $36.2^{+0.7}_{-0.6}$ & $3327\pm 58$ & $>474$ &     ...     & $565\pm 19$ & $730\pm 37$ & $1129\pm 24$ \\

SDSSJ\,095000.86$+$483102.2 & 0.2119 &  93.7 & & 0.21169 & $-52$ & all& ... & $18.52^{+0.05}_{-0.10}$ & ...& $1353\pm 27$ & $689\pm 15$ & $206\pm 14$ & $500\pm 28$ & $312\pm 16$ & $584\pm 12$ \\
 &  &   & &&&1&  $-96.9^{+9.5}_{-3.1}$ & $15.43^{+0.13}_{-0.05}$ & $25.8^{+6.2}_{-2.5}$& ... & ... & ... & ... &    ...  &  ... \\
 &  &   & &&&2&  $0.0^{+0.7}_{-0.5}$ & $18.52^{+0.05}_{-0.10}$ & $24.7^{+0.8}_{-1.0}$ & ... & ... & ... & ... &    ...  &  ... \\
 &  &   & &&&3&  $+95.9^{+2.6}_{-5.0}$ & $15.06^{+0.11}_{-0.07}$ & $14.4^{+3.4}_{-1.7}$ & ... & ... & ... & ... &    ...  &  ... \\

SDSSJ\,112755.83$+$115438.3 & 0.4237 &  98.7 & & 0.42461 & $+192$ & all &  ... & $15.81\pm0.06$ & ... &  $680\pm 31$ & $185\pm 10$ & $54\pm 12$  & $90\pm 33$  &     ...     & $62\pm 26$ \\
 &  &   & &&&1&  $-64.4^{+4.6}_{-3.6}$ & $15.45^{+0.07}_{-0.06}$ & $29.0^{+3.3}_{-2.5}$ & ... & ... & ... & ... &    ...  &  ... \\
 &  &   & &&&2&  $0.0^{+3.5}_{-3.4}$ & $15.56^{+0.07}_{-0.11}$ & $16.9^{+2.4}_{-2.3}$ & ... & ... & ... & ... &    ...  &  ... \\

SDSSJ\,124307.36$+$353926.3 & 0.3896 & 102.2 & & 0.3896 & ... & ... &   $0.0$ &    $<\,12.8$    & $15^a$ &    $<\,19$ &     $<\,12$ &      $<\,8$ &     $<\,18$ &    $<\,34$  & ... \\

SDSSJ\,155047.70$+$400122.6 & 0.3125 & 106.7 & & 0.31282 & $+73$ & all&  ...  & $16.61\pm0.04$ & ... & $1264\pm 24$ & $445\pm 27$ & $72\pm 23$  & $360\pm 26$ &    $<\,105$  & $196\pm 11$ \\
&  &   & &&&1&  $-57.1^{+3.3}_{-3.0}$ & $16.21\pm0.06$ & $39.1^{+1.8}_{-1.7}$& ... & ... & ... & ... &    ...  &  ... \\
&  &   & &&&2&  $0.0\pm0.1$ & $16.38^{+0.07}_{-0.08}$ & $13.1\pm1.0$ & ... & ... & ... & ... &    ...  &  ... \\
&  &   & &&&3&  $+83.2^{+4.8}_{-7.3}$ & $14.58\pm0.05$ & $54.1^{+7.1}_{-4.3}$ & ... & ... & ... & ... &    ...  &  ... \\
%SDSSJ\,155048.29$+$400144.9 & 0.3125 & 106.7 & & 0.3126 & &  $22.8$ & $16.6\pm 0.1$ & 25.0 & $1264\pm 24$ & $445\pm 27$ & $72\pm 23$  & $360\pm 26$ &    $<\,36$  & $196\pm 11$ \\

SDSSJ\,024651.20$-$005914.1 & 0.4105 & 108.6 & & 0.41168 & $+251$ & all & ... & $17.21\pm 0.01$ & ... & $1573\pm 34$ & $365\pm 26$ & $45\pm 8$   & $144\pm 42$  &   $146\pm 32$ & $84\pm 15$ \\
 &  &   & &&&1&  $-310.6^{+6.5}_{-5.5}$ & $13.96\pm0.06$ & $59.0^{+3.1}_{-2.5}$& ... & ... & ... & ... &    ...  &  ... \\
 &  &   & &&&2&  $-229.8^{+0.8}_{-0.7}$ & $16.49^{+0.06}_{-0.05}$ & $17.4\pm0.3$ & ... & ... & ... & ... &    ...  &  ... \\
 &  &   & &&&3&  $-128.5^{+0.7}_{-0.9}$ & $14.04\pm0.05$ & $37.2^{+1.1}_{-1.4}$ & ... & ... & ... & ... &    ...  &  ... \\
 &  &   & &&&4&  $+0.0^{+0.9}_{-0.8}$ & $17.11^{+0.01}_{-0.02}$ & $16.6\pm0.4$ & ... & ... & ... & ... &    ...  &  ... \\
 &  &   & &&&5&  $+52.5^{+2.5}_{-2.0}$ & $15.62^{+0.07}_{-0.08}$ & $24.2^{+0.8}_{-0.9}$ & ... & ... & ... & ... &    ...  &  ... \\

SDSSJ\,135727.27$+$043603.3 & 0.3296 & 125.9 & & 0.32869 & $-205$ & 1 & $0.0\pm0.4$ & $17.483\pm 0.002$ & $18.3\pm0.2$ &  $686\pm 34$ &     $<\,24$ &   $<\,22$   &     $<\,46$ &     $<\,47$ & $<\,54$ \\

SDSSJ\,091027.70$+$101357.2 & 0.2641 & 140.1 & & 0.26341 & $-164$ & all & ... & $16.65^{+0.34}_{-0.24}$ & ... & $1198\pm 38$ & $681\pm 48$$^c$ &     ...     & $214\pm 60$ &  $73\pm 30$ & $145\pm 18$ \\
 &  &   & &&&1&  $0.0\pm0.7$ & $16.63^{+0.35}_{-0.25}$ & $14.9^{+1.0}_{-1.5}$ & ... & ... & ... & ... &    ...  &  ... \\
 &  &   & &&&2&  $+91.5^{+2.9}_{-2.8}$ & $14.34^{+0.15}_{-0.16}$ & $19.2^{+2.6}_{-2.7}$& ... & ... & ... & ... &    ...  &  ... \\
 &  &   & &&&3&  $+218.7^{+2.3}_{-2.8}$ & $15.11^{+0.09}_{-0.10}$ & $40.5^{+2.9}_{-2.1}$ & ... & ... & ... & ... &    ...  &  ... \\

SDSSJ\,141307.39$+$091956.7 & 0.3584 & 149.2 & & 0.3584 & ... & ...  &   $0.0$ &    $<\,12.8$  & $15^a$ &    $<\,25$   &      $<\,9$ &   $<\,6$    &     $<\,12$ &     $<\,17$ & ... \\

SDSSJ\,155304.32$+$354853.9 & 0.4736 & 155.9 & & 0.47540 & $+366$ & 1 & $0.0^{+1.1}_{-1.2}$ & $15.64\pm0.04$ & $22.9^{+1.2}_{-1.0}$ &  $638\pm 76$ & $206\pm 12$ & $75\pm 25$  &    $<\,119$ &    ...      & $49\pm 8$ \\

SDSSJ\,125859.98$+$413128.2 & 0.2790 & 159.1 & & 0.2790 & ... & ...   &  $0.0$ &    $<\,12.8$    & $15^a$ &    $<\,15$     & $<\,13$ &   $<\,14$   &     $<\,14$ &     $<\,21$ & ... \\

SDSSJ\,124409.17$+$172111.9 & 0.5591 & 160.1 & & 0.55851 & $-113$ & 1 &$0.0^{+1.3}_{-1.4}$ & $15.98\pm 0.03$ & $30.2^{+1.9}_{-1.7}$ &    $416^b$   & $306\pm 14$ &     ...     &     ...     &    ...      & $130\pm 5$ \\

\hline 
\multicolumn{16}{l}{$^\mathrm{a}$The purpose of specifying the
  $b$ value here is for generating the Voigt profile displayed in
  Figures 2 \& 3 for the non-detections.} \\ 
\multicolumn{16}{l}{$^\mathrm{b}$\lya\ is not covered by the
  available COS spectrum; \ewlya is inferred from the best-fit \lnhone
  and $b$.} \\ 
\multicolumn{16}{l}{$^\mathrm{c}$Werk \etal\ (2013) reported
  $W_r(977)=282\pm 63$ m\AA\ for this feature.  The discrepancy is
  likely due to a combination of low $S/N$ and uncertainties in the
  continuum normalization.} \\
\end{tabular}
}
\end{table}
\end{landscape}

\subsection{Optically-thick gas in LRG Halos}

Our survey shows not only that \HI\ absorbers are frequently seen at
$d<160$ kpc in LRG halos, but that a large fraction (7/16) of these
absorbers are also optically thick to ionizing radiation field with
$\log\,N(\HI)>17.2$, corresponding to an optical depth at the Lyman
edge of $\tau_{\rm 912}>1$.  Such a high rate of incidence exceeds the
expectation from random background counts.  We estimate the expected
number of random Lyman limit systems (LLS) in the redshift range
surveyed in the COS-LRG sample by adopting the number density of $z<1$
LLS from Ribaudo \etal\ (2011a) and Shull \etal\ (2017).  Over the
search window of $\pm 500$ \kms\ around 16 LRGs at $z\approx 0.4$, we
only expect to find between 0.02 and 0.04 LLS over the entire COS-LRG
sample.

In addition, the observed large clustering amplitude of LRGs (e.g.,
Zheng \etal\ 2007; Padmanabhan \etal\ 2007; Gauthier \etal\ 2009) also
indicates that there are on average more lower-mass galaxies clustered
around the LRGs than typical $L_*$ galaxies on scales of Mpc and,
through redshift space distortions, absorbing gas in these correlated
halos could contribute to the detection statistics around LRGs.  No
clustering measurements are available for LLS at low redshifts.  We
therefore infer the contribution from correlated galaxy halos from
observations of \MgII\ in the outskirts of galaxy clusters (Lopez
\etal\ 2008; see also Muzahid \etal\ 2017).  Lopez \etal\ (2008)
reported a factor of $4-9$ overabundances in strong \MgII\ absorbers
($\ewmgii>1$ \AA) at $d<1$ Mpc from galaxy clusters.  Taking into
account possible enhancement due to clustering leads to a total of
between 0.08 and 0.36 expected random and correlated LLS over the $\pm
500$ \kms\ velocity window along 16 LRG sightlines.  This relatively
small number of expected LLS from random and correlated counts, in
contrast to seven LLS detected at $d<160$ kpc and $|v_{{\rm gas}-{\rm
    gal}}|\apl 350$ \kms\ from 16 LRGs, provides a strong support for
a physical connection between the observed optically-thick gas and the
LRGs.

We present the spatial distribution of $N(\HI)$ as a function of $d$
for the COS-LRG sample in Figure 4a.  For comparison, we also include
the COS-Halos red galaxies (Thom \etal\ 2012) with updated $N(\HI)$
from Prochaska \etal\ (2017).  In Figure 4b, we show the distribution
of $N(\HI)$ versus \mstar\ for both samples.  No correlation is found
between $N(\HI)$ and $M_{\rm star}$.  Recall that COS-LRG sample is
uniformly defined to have $\log\,\mstar\,/\,\msun\apg 11$ with a
median and dispersion of
$\langle\,\log\,\mstar\,/\,\msun\,\rangle_{\rm COS-LRG}=11.2$ and
$\sigma_{\langle\,\log\,\mstar\,/\,\msun\,\rangle_{\rm COS-LRG}}=0.2$
dex at $z=0.21-0.55$, while the COS-Halos red sample includes a large
fraction of lower-mass galaxies with a median and dispersion of
$\langle\,\log\,\mstar\,/\,\msun\,\rangle_{\rm COS-Halos Red}=10.8\pm
0.3$ and $\sigma_{\langle\,\log\,\mstar\,/\,\msun\,\rangle_{\rm
    COS-Halos Red}}= 0.3$ dex at $z=0.14-0.27$.  While there exists a
substantial scatter in the observed $N(\HI)$, a large fraction of the
COS-LRG and the COS-Halos red samples contain optically-thick gas at
$d\apl 100$ kpc and a general declining trend of $N(\HI)$ is also seen
toward larger distances.

\begin{figure}
\includegraphics[scale=0.33]{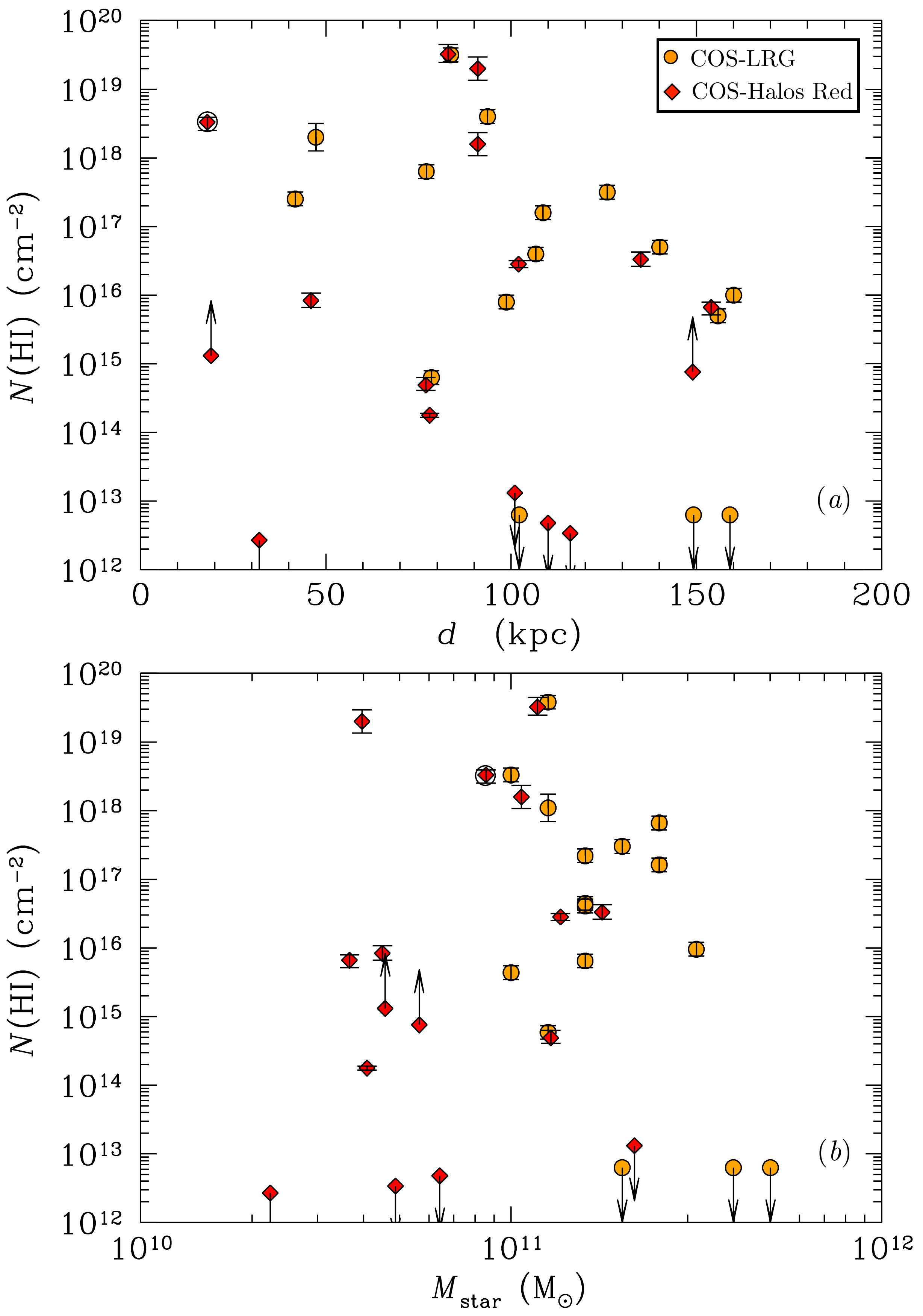}
\caption{{\it Top}: The observed spatial distribution of $N(\HI)$ as a
  function of projected distance $d$ for the COS-LRG sample (filled
  circles), in comparison to those found for the COS-Halos red
  galaxies (red diamonds; Thom \etal\ 2012).  Downward arrows
  represents a 2-$\sigma$ upper limit to the underlying $N(\HI)$ for
  halos with no \lya\ absorption line detected.  The COS-Halos
  measurements have been updated based on the report of Prochaska
  \etal\ (2017).  For two of the COS-Halos red galaxies, the available
  Lyman series lines are all saturated and therefore only lower limits
  to $N(\HI)$ are available.  These are shown as upward arrows.  {\it
    Bottom}: Distribution of $N(\HI)$ versus $M_{\rm star}$ for both
  COS-LRG and COS-Halos red samples.  Recall that COS LRGs form a
  mass-limited sample of $\log\,\mstar\apg 11$ with a median of
  $\langle\,\log\,\mstar\,/\,\msun\,\rangle_{\rm COS-LRG}=11.2$ at
  $z=0.2-0.55$, while the COS-Halos red sample includes predominantly
  lower-mass galaxies with a median of
  $\langle\,\log\,\mstar\,/\,\msun\,\rangle_{\rm COS-Halos Red}=10.8$
  at $z=0.14-0.27$.  No correlation is found between $N(\HI)$ and
  $M_{\rm star}$.}
        \label{figure:spatial1}
\end{figure}

We determine the covering fraction and associated uncertainties of
optically-thick gas in massive quiescent halos following the
maximum-likelihood approach described in Chen \etal\ (2010).  The
likelihood of observing an ensemble of LRGs with $n$ showing
associated LLS and $m$ non-detections is given by
\begin{equation}
  \begin{split}
  \mathcal{L}(\kappa_{\rm LLS}) = \langle\kappa\rangle_{\rm LLS}^n
          [1-\langle\kappa\rangle_{\rm LLS}]^m.
  \end{split}
\end{equation}
The full COS-LRG sample probes the LRG halos out to $d\approx 160$ kpc
(or $d\approx 1/3\,R_{\rm vir}$).  Based on the detections of LLS in
seven of the 16 LRG halos, we estimate a mean covering fraction of
optically-thick gas of $\langle\kappa\rangle_{\rm
  LLS}=0.44^{+0.12}_{-0.11}$ at $d\apl 160$ kpc.  The error bars
represent the 68\% confidence interval.  In the inner halos, where we
detect LLS in five of the seven LRG halos, the mean covering fraction
is $\langle\kappa\rangle_{\rm LLS}=0.71^{+0.11}_{-0.20}$ at $d\apl
100$ kpc.  We note that excluding five COS-Halos red galaxies does not
change the results.  We estimate $\langle\kappa\rangle_{\rm
  LLS}=0.45^{+0.14}_{-0.13}$ at $d\apl 160$ kpc based on five
detections of LLS around 11 randomly selected LRGs.

The observed covering fraction of optically-thick gas in massive
quiescent halos at $z\approx 0.4$ is surprisingly consistent with the
measurement of $\langle\kappa\rangle_{\rm LLS}=0.6^{+0.2}_{-0.1}$ at
$d\apl 80$ kpc from $L_*$ and sub-$L_*$ star-forming galaxies at
$z\approx 0.2$ from the COS-Halos survey (e.g., Prochaska
\etal\ 2017).  It is also very comparable to
$\langle\kappa\rangle_{\rm LLS}=0.64^{+0.07}_{-0.10}$ observed at
$d\apl 122$ kpc in quasar host halos (e.g., Hennawi \etal\ 2006;
Prochaska \etal\ 2013) and $\langle\kappa\rangle_{\rm LLS}=0.30\pm
0.14$ observed at $d\apl R_{\rm vir}$ in halos hosting starburst
galaxies at $z\approx 2.3$ (e.g., Rudie \etal\ 2012).  Quasars are
typically found to reside in dark matter halos of mass $M_h\sim
10^{12.5}\ {\rm M}_\odot$ (e.g., White et al.\ 2012), while the
targeted starburst galaxies at $z\approx 2.3$ have a characteristic
halo mass of $M_h\sim 10^{12}\ {\rm M}_\odot$ (e.g., Adelberger
\etal\ 2005).  Both populations reside in lower-mass halos at $z>2$
than the LRGs at $z\approx 0.4$.  To reproduce the observed high gas
covering fraction in these distant halos in cosmological simulations,
it is necessary to invoke energetic feedback due to either active
galactic nuclei (AGN) in quasar host galaxies (e.g., Rahmati
\etal\ 2015) or massive stars (e.g., Faucher-Gigu\`ere \etal\ (2016).
The quiescent state of the LRGs makes either AGN or stellar feedback a
challenging scenario for explaining the frequent appearance of
optically-thick gas in these massive halos.

At the same time, we note the stark contrast between LRG halos at
$z\approx 0.4$ and the M31 halo.  For the M31 halo, the covering
fraction of optically-thick gas is constrained to be $<5$\% based on
deep 21~cm surveys (Howk \etal\ 2017), while the covering fraction of
\SiIII\ absorbing gas is found to approach unity from absorption
spectroscopy of background QSOs (Lehner \etal\ 2015).  The lack of
optically-thick gas coupled with a high covering fraction of heavy
ions suggests that either the halo gas is highly ionized or the
absorbing clouds are significantly smaller than the beam size
($\approx 2$ kpc in diameter) of the 21~cm observations (e.g., Rigby
\etal\ 2002; Stern \etal\ 2016).  M31 with a star formation rate (SFR)
of $\approx 0.4\,{\rm M}_\odot\ {\rm yr}^{-1}$ (e.g., Barmby
\etal\ 2006; Rahmani \etal\ 2016) resides in a dark matter halo of
$M_h\approx 10^{12}\,{\rm M}_\odot$ (e.g., Watkins \etal\ 2010; Sofue
2015).  A lack of optically-thick gas would be qualitatively
consistent with the observed low SFR in M31 disk, but different from
what is observed for massive quiescent halos at $z\approx 0.4$.

With the observed large covering fraction of optically-thick gas in
LRG halos, we proceed to evaluate the fractional contribution of LRGs
to the LLS observed along random QSO sightlines.  The expected number
of LLS per unit redshift per line of sight originating in LRG halos is
computed following
\begin{align}
n_{\rm LLS}({\rm LRG})=\frac{c}{H_0}\frac{(1+z)^2}{\sqrt{\Omega_M(1+z)^3+\Omega_\Lambda}}\times \nonumber \\
\int_{L_{\rm min}}^\infty dL\,\phi(L)\langle\kappa\rangle_{\rm LLS}(\pi R_{\rm gas}^2),
\end{align}
where $c$ is the speed of light, $\phi(L)$ is the LRG luminosity
function, $R_{\rm gas}$ is the radius of gaseous halos, and $L_{\rm
  min}$ is the minimum luminosity of LRGs which is approximately
$3\,L_*$ for the COS-LRG sample, with a corresponding absolute
$i$-band magnitude of $M_i=-22.6$.  Here we have assumed that the gas
cross section does not vary significantly with luminosity.  Adopting
the LRG luminosity function of Montero-Dorta \etal\ (2016), which is
characterized by $M_{i*}=-21.4$ and $\phi_*=7.58\times 10^{-4}\ {\rm
  Mpc}^{-3}$, we find $n_{\rm LLS}({\rm LRG})\approx 0.0053$ for
$R_{\rm gas}=160$ kpc and $\langle\kappa\rangle_{\rm LLS}=0.44$.  The
observed number density of LLS with $\log\,N(\HI)\ge 17.2$ is $n_{\rm
  LLS}=0.36-0.5$ at $z<1$ from Ribaudo \etal\ (2011a) and Shull
\etal\ (2017).  We therefore conclude that despite a high covering
fraction of optically-thick gas in the inner 160 kpc in radius of
massive quiescent halos, these objects are too rare to contribute
more than 2\% of the observed LLS at $z<1$ along random QSO
sightlines.

\subsection{Incidence of Heavy Ions around LRGs}

A primary motivation of the COS-LRG program is to understand the
origin and significance of \MgII\ absorbers found in LRG halos (e.g.,
Gauthier \etal\ 2009, 2010, Gauthier \& Chen 2011; Bowen \& Chelouche
2011; Huang \etal\ 2016).  The COS-LRG sample is established without
prior knowledge of the presence/absence of any absorption features in
the halos, including metal absorption lines.  It therefore provides a
necessary baseline calibration for characterizing the significance of
metal-line absorbers in these massive quiescent halos.  The expanded
spectral coverage from combining COS FUV and ground-based echelle
spectra also provides empirical constraints for multiple ionization
species, from singly-ionized silicon and magnesium, to twice-ionized
carbon and silicon, and to highly-ionized O$^{5+}$.

\begin{figure}
\includegraphics[scale=0.375]{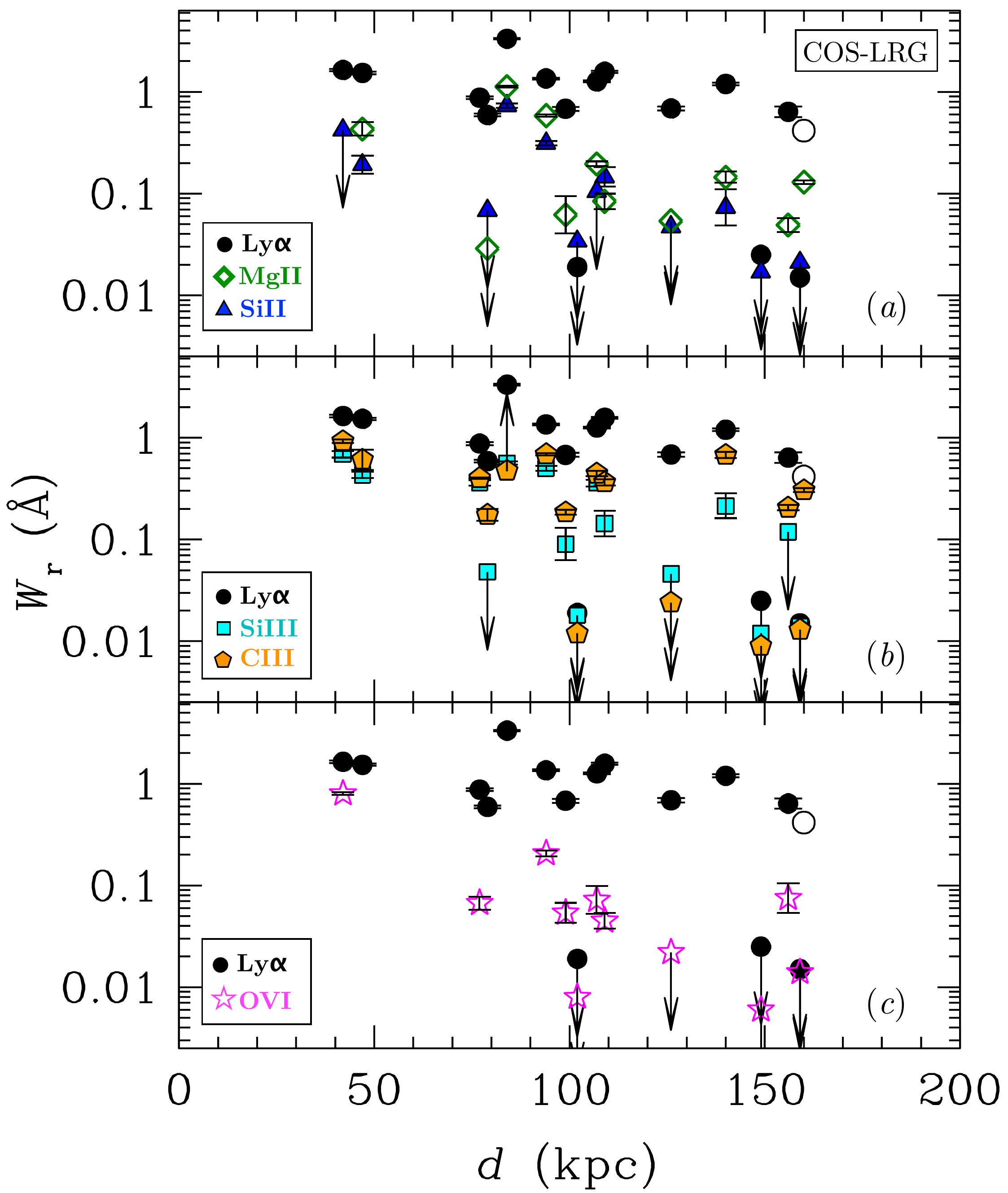}
\caption{Measurements of rest-frame absorption equivalent width, $W_r$
  versus $d$ for different absorption transitions.  Panel ({\it a})
  presents low-ionization $\MgII\,\lambda\,2796$ (open diamonds) and
  $\SiII\,\lambda\,\,1260$ (filled triangles) transitions; panel ({\it
    b}) presents intermediate-ionization $\SiIII\,\lambda\,1206$
  (filled squares) and $\CIII\,\lambda\,977$ (filled pentagons)
  transitions; and panel ({\it c}) presents high-ionization
  $\OVI\,\lambda\,1031$ transitions (star symbols).  Measurements for
  \lya\ (black circles) are included in all three panels for
  comparisons.  For SDSSJ\,124409.17$+$172111.9 at $z=0.5591$ and
  $d=160$ kpc, \ewlya is shown as an open circle because \lya\ is not
  covered by the available COS spectrum and \ewlya is inferred from
  the best-fit \lnhone and $b$.  For all but one LRG halos with
  \lya\ detected, associated metal absorption lines are also detected
  with $\CIII\,\lambda\,977$ and $\SiIII\,\lambda\,1206$ being the two
  most dominant features.  The only exception is
  SDSSJ\,135726.27$+$043541.4 at $z=0.3296$ and $d=126$ kpc for which
  optically-thick gas is present with $\log\,N(\HI)=17.5$ but no metal
  absorption lines are detected to 2-$\sigma$ limits of $<0.05$ \AA.}
        \label{figure:spatial2}
\end{figure}

Figure 5 shows the rest-frame absorption equivalent width, $W_r$
versus $d$ for \lya\ and associated metal absorption features,
including $\MgII\,\lambda\,2796$, $\SiII\,\lambda\,\,1260$,
$\SiIII\,\lambda\,1206$, $\CIII\,\lambda\,977$, and
$\OVI\,\lambda\,1031$.  For all but one LRG halos with \HI\ absorbers
detected, associated metal absorption features are also detected with
intermediate ions revealed by $\CIII\,\lambda\,977$ and
$\SiIII\,\lambda\,1206$ being the most dominant ionization state,
exceeding both low-ionization gas probed by \MgII\ and highly-ionized
gas probed by \OVI.  For the three LRGs without an \HI\ absorber
detected, no metal absorption lines are found to sensitive upper
limits.

SDSSJ\,135726.27$+$043541.4 at $z=0.3296$ and $d=126$ kpc is the only
LRG displaying no trace of metal absorption lines to 2-$\sigma$ limits
of $<0.09$ \AA, while showing a strong Lyman absorption series with
$\log\,N(\HI)=17.5$.  The presence of optically-thick gas combined
with a complete absence of ionic absorption features places a
stringent upper limit of gas metallicity at $\apl 0.3$\% solar (see
Zahedy \etal\ 2018 in preparation), representing the
lowest-metallicity gas detected near a luminous galaxy at intermediate
redshift (cf.\ Ribaudo \etal\ 2011b; Prochaska \etal\ 2017).  This low
metallicity is among the most metal-poor LLS at $z<1$ (e.g., Wotta
\etal\ 2016) and suggests that the absorber originates in inflows of
pristine IGM into the LRG halo (e.g., Hafen \etal\ 2017).  The
detection of a chemically-pristine Lyman limit absorber at $z=0.33$
also shows that intergalactic medium at $z<1$ may not be as
chemically-enriched as previously thought (e.g., Shull \etal\ 2014;
Lehner 2017).

\begin{figure}
\includegraphics[scale=0.45]{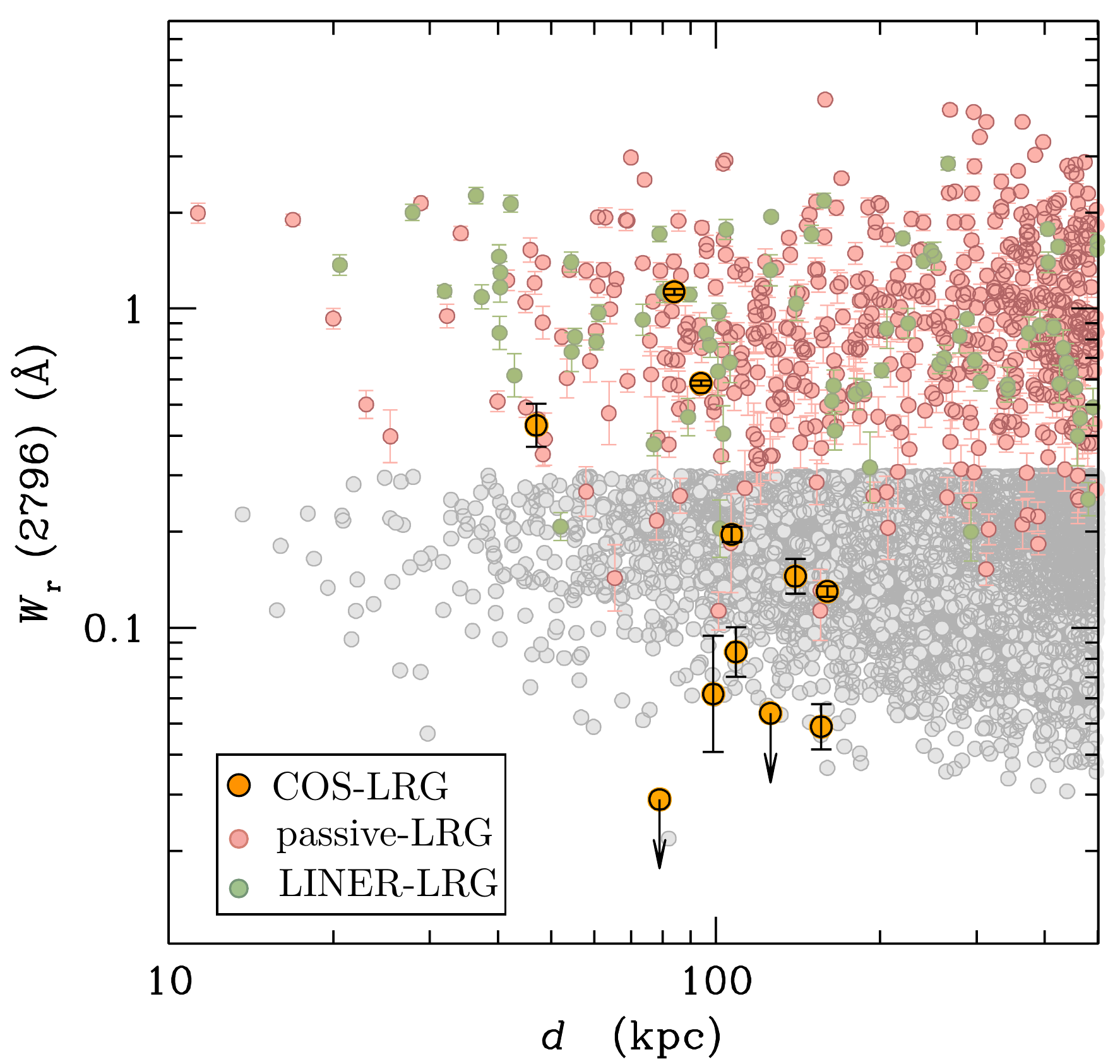}
\caption{Comparison of rest-frame absorption equivalent width,
  $\ewmgii$ versus $d$ for 11 COS LRGs (orange circles) with
  ground-based echelle absorption spectra available and the SDSS LRG
  samples from Huang \etal\ (2016).  Recall that roughly 10\% of LRGs
  exhibit LINER features (e.g., Johnston \etal\ 2008; Hodge
  \etal\ 2008).  These LINER-like galaxies are shown in green circles,
  while the remaining 90\% of passive LRGs are shown in red.
  Available echelle spectra for the COS LRGs have allowed us to
  uncover very weak \MgII\ absorption features of total integrated
  $\ewmgii \apg 0.05$ \AA\ that have been previously missed in SDSS
  spectra.  While only three of the 11 COS LRGs have associated
  \MgII\ absorbers of $\ewmgii > 0.3$ \AA, only two of these galaxies
  without associated \MgII\ absorption detected to a 2-$\sigma$ upper
  limit of $\ewmgii\approx 50$ m\AA.  }
        \label{figure:comparison}
\end{figure}

The presence of heavy ions in all but one COS LRG halo with strong
\HI\ absorbers indicates a wide spread chemical enrichment in these
massive quiescent halos.  An early result of the COS-Halos survey is
the \OVI\ ``bimodality'' in galaxy halos (Tumlinson \etal\ 2011):
galaxies with higher specific-SFR exhibit on average stronger
\OVI\ absorption features in their halos than those with low
specific-SFR.  This phenomenon is commonly attributed to a causal
connection between star formation in galaxies and stellar-feedback
driven chemical enrichment in galaxy halos (e.g., Ford \etal\ 2013;
Suresh \etal\ 2015), in which case the observed heavy elements in
galaxy halos are ejected from galaxies by newly formed young stars and
the origin of heavy elements in quiescent halos remains unknown.
Alternatively, because low specific-SFR galaxies in the COS-Halos
sample are also more massive than their star-forming counterparts, the
deficit of strong \OVI\ absorption observed in these passive galaxies
can also be understood as due to oxygen being ionized to higher
ionization states (e.g., Oppenheimer \etal\ 2016).  In this case, the
presence/absence of \OVI\ is not directly connected to star formation
or AGN feedback, but it requires the CGM to be pre-enriched to $\apg
1/10$ solar at $<R_{\rm vir}$.  Adopting an equivalent width threshold
of 0.1\,\AA, we estimate a mean gas covering fraction of
$\langle\kappa(\OVI)\rangle_{\rm 0.1}=0.18_{-0.06}^{+0.16}$,
consistent with the finding for COS-Halos red galaxies (e.g., Werk
\etal\ 2013).  While quiescent galaxies exhibit on average fewer
strong \OVI\ absorbers of column density $\log\,N(\OVI)>14$ than
star-forming galaxies at $d<160$ kpc, the mean gas covering fraction
of moderately strong \OVI\ absorbers of $\log\,N(\OVI)>13.5$ is found
to remain high at $\approx 50$\% in quiescent halos (Johnson
\etal\ 2015b).

Of the 16 galaxies in the COS-LRG sample, five do not have constraints
for \OVI\ absorption because the spectral regions are contaminated by
either geocoronal emission or other strong features.  For the 11 LRGs
with available constraints on the \OVI\ absorption strength, six have
associated \OVI\ absorbers of $\ewovi>0.05$ \AA, corresponding roughly
to $\log\,N(\OVI)>13.5$ under the linear curve-of-growth assumption.
We estimate a mean covering fraction of
$\langle\kappa(\OVI)\rangle_{\rm 0.05}=0.55_{-0.14}^{+0.13}$,
extending previous findings to higher-mass quiescent galaxies.

Similarly, 11 of the 16 QSO-LRG pairs have optical echelle spectra of
the QSOs available for constraining the absorption strengths of the
associated \MgII\ lines.  These high signal-to-noise ($S/N$) and high
spectral resolution echelle spectra have allowed us both to uncover
very weak \MgII\ absorption features of total integrated $\ewmgii \apg
0.05$ \AA\ that have been previously missed in SDSS spectra and to
resolve individual components (see Figure 3).  While only three of the
11 COS LRGs have associated \MgII\ absorbers of $\ewmgii > 0.3$ \AA,
only two of these LRGs do not have associated \MgII\ absorption
detected to a 2-$\sigma$ upper limit of $\ewmgii\approx 50$ m\AA.
Figure 6 presents a comparison of \ewmgii\ versus $d$ between the
COS-LRG sample and the SDSS LRG samples, both passive (red symbols)
and LINER-like (green), from Huang \etal\ (2016).

The mean covering fraction of \MgII\ absorbing gas at $d\apl 160$ kpc
is found to be $\langle\kappa(\MgII)\rangle_{\rm
  0.3}=0.27^{+0.16}_{-0.09}$ for a minimum absorption equivalent width
$W_0=0.3$ \AA, consistent with the mean covering fraction found using
the full SDSS LRG sample from Huang \etal\ (2016).  For a more
sensitive limit of $W_0=0.1$ \AA, we estimate
$\langle\kappa(\MgII)\rangle_{\rm 0.1}=0.55^{+0.13}_{-0.14}$ at
$d<160$ kpc, comparable to what is observed in lower-mass,
star-forming galaxies (e.g., Chen \etal\ 2010; Werk \etal\ 2013).  As
a cautionary note, none of the three ``transparent'' halos with no
detectable \lya\ have optical echelle spectra available.
Consequently, these LRGs do not contribute to the estimates of
\MgII\ gas covering fraction here.  The estimated mean value above
therefore likely represents an upper limit to the true value.
Including the three ``transparent'' halos as non-detections in
\MgII\ places a lower limit at $\langle\kappa(\MgII)\rangle_{\rm
  0.1}>0.43$.  Combining these limits, we therefore infer a mean
covering fraction of \MgII\ absorbing gas in the range of
$\langle\kappa(\MgII)\rangle_{\rm 0.1}\approx 0.4-0.55$ in LRG halos.
%We
%consider this a lower limit because the remaining two missing
%sightlines without available optical echelle spectra
%(SDSSJ\,094632.40$+$512335.9 at $z=0.407$ and $d=42$ kpc, and
%SDSSJ\,111132.33$+$554712.8 at $z=0.4629$ and $d=77$ kpc) both exhibit
%strong metal-line transitions at the locations of the LRGs in the COS
%data.  Therefore, these   

A surprising finding of the COS-LRG survey is the significantly higher
incidence of \CIII\ than \MgII\ absorption features around LRGs,
indicating that chemically-enriched cool gas is more abundant in these
massive quiescent halos than previously thought from searches of
associated \MgII\ absorption.  Twelve of the 16 LRGs in the sample
exhibit associated \CIII\ with $\ewciii > 0.1$ \AA, leading to a mean
covering fraction of $\langle\kappa(\CIII)\rangle_{\rm
  0.1}=0.75^{+0.08}_{-0.13}$ for a minimum rest-frame absorption
equivalent width of $W_0=0.1$ \AA\ at $d\apl 160$ kpc.  The observed
high covering fraction of \CIII\ absorbing gas in LRG halos is
comparable with what is reported for COS-Halos red galaxies with
$\langle\kappa(\CIII)_{\rm COS-Halos red}\rangle_{\rm
  0.1}=0.79^{+0.11}_{-0.31}$ by Werk \etal\ (2013), which is also
statistically consistent with what is seen in COS-Halos blue galaxies
with $\langle\kappa(\CIII)_{\rm COS-Halos blue}\rangle_{\rm
  0.1}=0.90\pm 0.07$ in the COS-Halos survey.  While many of the
detected \CIII\ lines are saturated and therefore only lower limits to
the underlying ionic column density $N(\CIII)$ can be obtained, \CIII,
unlike \OVI, does not appear to show a strong bimodality in galaxy
halos.

Similarly, \SiIII\,$\lambda$\,1206 also appears to be dominant,
although available COS spectra do not provide comparable sensitivities
for placing as strong constraints for \ewsiiii\ as for
\CIII\,$\lambda$\,977.  Adopting a minimum rest-frame absorption
equivalent width of $W_0=0.1$ \AA\ for \SiIII, we conservatively
estimate a mean gas covering fraction of
$\langle\kappa(\SiIII)\rangle_{\rm 0.1}=0.4-0.7$, similar to what was
found for COS-Halos red galaxies (e.g., Werk \etal\ 2013).

\subsection{Implications}

The dominant presence of intermediate ions revealed by \CIII\ and
\SiIII\ absorption constrains the temperature of the absorbing gas to
be $T\apl 3\times 10^4$ K under either collisional or photo-ionization
scenarios (see e.g., Oppenheimer \& Schaye 2013).  This is in contrast
to $T\apg 10^{6.5}$ K expected for the hot halo as a result of
gravitational heating, and supports a multi-phase halo model, in which
cool absorbing clouds are pressure-confined in a hot medium.  Under a
pressure equilibrium, this would imply a density contrast between cool
clouds and the hot halo of $\sim 100:1$.  While dense, cool clouds are
expected to fall in a low-density hot halo, whether or not they can
reach the galaxy to fuel star formation depends on the infall time
relative to the disruption time (e.g., Maller \& Bullock 2004).  Under
a simple hydrostatic equilibrium assumption, Gauthier \& Chen (2011)
calculated the minimum mass required for cool clouds to survive
thermal conduction from the surrounding hot medium.  They found a
minimum mass of $M_{\rm cl}\apg 10^6\,\msun$.  On the other hand,
Huang \etal\ (2016) reported that the velocity dispersion of Mg\,II
absorbing gas relative to the host LRGs is suppressed in comparison to
the expectations from virial motion.  These authors attributed the
suppressed velocity dispersion to ram-pressure drag force and inferred
a maximum cloud mass of $M_{\rm cl}\approx 5\times 10^4\,\msun$.
While combining these calculations suggests that the majority of cool
clouds formed at large distances may not be sufficiently massive to
survive the hot halo during infall, we note that the results of these
calculations depend sensitively on the model assumptions of the
properties of the hot halo and how cool clouds form (e.g., Brighenti
\& Mathews 2003).  Recall that high column density gas is detected at
$d<20$ kpc around massive lensing galaxies at $z\approx 0.4$ (e.g.,
Zahedy \etal\ 2016, 2017a) and in nearby ellipticals (e.g., Oosterloo
\etal\ 2010; Serra \etal\ 2012; Young \etal\ 2014).

We note that while our study focuses entirely on LRGs selected by
their quiescent state, the results bear significantly on our overall
understanding of feedback mechanisms in galaxy halos.  More than 90\%
of massive galaxies with $\mstar \apg 10^{11}\,\msun$ in the local
universe are observed to contain primarily evolved stellar populations
with little on-going star formation (e.g., Peng et al. 2010; Tinker
\etal\ 2013).  The results from our study can therefore be applied
broadly to massive halos of $\mstar \apg 10^{11}\,\msun$, but in the
still higher-mass galaxy cluster regime observations have yielded very
different results (see e.g., Yoon \etal\ 2012; Burchett \etal\ 2018).
In addition, the LRGs also provide a unique sample for identifying
additional physical mechanisms, beyond starburst and AGN feedback,
that produce the observed optically-thick clouds and heavy ions in
galactic halos.  A joint study of the ionization state of the gas and
the galaxy environment is the next important step for obtaining key
insights into these underlying feedback mechanisms.

\section{Summary and Conclusions}

We have carried out an absorption-line survey of cool gas in halos
around a mass-limited sample of LRGs with $\log\,\mstar\ge 11$ at
$z=0.21-0.55$.  The LRGs are selected with no prior knowledge of the
presence/absence of any absorption features.  Our program is designed
such that accurate and precise measurements of $N(\HI)$ are possible
based on observations of the full hydrogen Lyman series.  Our analysis
shows that not only \HI\ absorbers are frequently seen at $d<160$ kpc
(or $d < 1/3\,R_{\rm vir}$) in LRG halos, but a large fraction (7/16)
of these absorbers are also optically thick to ionizing radiation
field with $\log\,N(\HI)>17.2$.  In addition, all but one LRGs with
detected \HI\ absorption also exhibit associated metal-line absorption
features, indicating a wide spread chemical enrichment in these
massive quiescent halos.  The main findings of our survey are
summarized below.

(1) The mean covering fraction of optically-thick gas is
$\langle\kappa\rangle_{\rm LLS}=0.44^{+0.12}_{-0.11}$ at $d\apl 160$
kpc and $\langle\kappa\rangle_{\rm LLS}=0.71^{+0.11}_{-0.20}$ at
$d\apl 100$ kpc in LRG halos.  These numbers are consistent with the
high covering fraction of $\langle\kappa\rangle_{\rm
  LLS}=0.6^{+0.2}_{-0.1}$ observed at $d\apl 80$ kpc from $L_*$ and
sub-$L_*$ star-forming galaxies at $z\approx 0.2$ from the COS-Halos
survey.

(2) Intermediate ions probed by \CIII\ and \SiIII\ are the most
prominent UV ionic features in these massive quiescent halos with a mean
covering fraction of
$\langle\kappa(\CIII)\rangle_{0.1}=0.75^{+0.08}_{-0.13}$ for
$\ewciii\ge 0.1$ \AA\ at $d<160$ kpc, comparable to what is seen for
\CIII\ in $L_*$ and sub-$L_*$ star-forming and red galaxies but
exceeding what is typically observed for strong \MgII\ or
\OVI\ absorbing gas around massive quiescent galaxies.

%Despite a high covering fraction of optically-thick gas in the
%inner 160 kpc in radius of massive quiescent halos, these objects are
%too rare to contribute more than 2\% of the observed LLS at $z<1$
%along random QSO sightlines.

(3) While quiescent galaxies exhibit on average fewer strong
\OVI\ absorbers of column density $\log\,N(\OVI)>14$ than star-forming
galaxies at $d<160$ kpc, the mean gas covering fraction of moderately
strong \OVI\ absorbers of $\log\,N(\OVI)>13.5$ is found to remain high
with a mean covering fraction of $\langle\kappa(\OVI)\rangle_{\rm
  0.05}=0.55_{-0.14}^{+0.13}$.

(4) The mean covering fraction of \MgII\ absorbing gas at $d\apl 160$
kpc is found to be $\langle\kappa(\MgII)\rangle_{\rm
  0.3}=0.27^{+0.16}_{-0.09}$ for $\ewmgii\ge 0.3$ \AA, consistent with
the mean covering fraction found using the full SDSS LRG sample.  For
absorbing gas of $\ewmgii\ge 0.1$ \AA, the mean covering fraction at
$d<160$ kpc is $\langle\kappa(\MgII)\rangle_{\rm
  0.1}=0.55^{+0.13}_{-0.14}$, comparable to what is observed in
lower-mass, star-forming galaxies.

In conclusion, the COS-LRG survey has uncovered a high incidence of
chemically-enriched cool ($T\sim 10^{4-5}$ K) gas in massive quiescent
halos hosting LRGs at $z\approx 0.4$.  A significant fraction of this
chemically-enriched gas is optically thick to ionizing photons.  It
shows that massive quiescent halos contain widespread
chemically-enriched cool gas.  No clear presence of bimodality is
found between blue star-forming and red quiescent galaxies in their
\HI\ gas content or in the incidence of intermediate ions.

\section*{Acknowledgments}

We thank Joop Schaye and Jonathan Stern for helpful discussions.  We
thank Michael Rauch for his help in obtaining the MIKE spectra for
some of the QSOs presented in this paper.  This work is based on data
gathered under the HST-GO-14145.01A program using the NASA/ESA Hubble
Space Telescope operated by the Space Telescope Science Institute and
the Association of Universities for Research in Astronomy, Inc., under
NASA contract NAS 5-26555.  HWC and FSZ acknowledge partial support
from NSF grant AST-1715692.  FSZ acknowledges generous support from
the Brinson Foundation and the Observatories of Carnegie Institution
for Science during his visit at Carnegie Observatories as a Brinson
Chicago--Carnegie predoctoral fellow.  SDJ is supported by a NASA
Hubble Fellowship (HST-HF2-51375.001-A).  This research has made use
of the Keck Observatory Archive (KOA), which is operated by the
W.~M.~Keck Observatory and the NASA Exoplanet Science Institute
(NExScI), under contract with the National Aeronautics and Space
Administration.

\label{lastpage}

\end{document}